\immediate\write18{makeindex -s nomencl.ist -o "\jobname.nls" "\jobname.nlo"}

\documentclass[journal,onecolumn,draftclsnofoot,]{elsarticle}

\usepackage{lineno,hyperref}
\modulolinenumbers[254]
\usepackage[ngerman,UKenglish]{babel}
\usepackage[top=60pt,bottom=65pt,left=60pt,right=55pt]{geometry}

\usepackage{multicol}
\usepackage{multirow}
\usepackage{arydshln}
\usepackage{lipsum}

\usepackage{setspace}
\usepackage{amsmath,bm}
\usepackage{fixmath}
\usepackage{commath}
\usepackage{lipsum}
\usepackage{textcomp}

\usepackage[bitstream-charter]{mathdesign}
\usepackage[T1]{fontenc}
\usepackage[utf8]{inputenc} 

\usepackage{framed}
\usepackage{nomencl} 
\makenomenclature

\biboptions{sort&compress} 

\usepackage{calrsfs}
\DeclareMathAlphabet{\pazocal}{OMS}{zplm}{m}{n}
\let\mathcal\undefined
\newcommand{\mathcal}[1]{\pazocal{#1}}

\usepackage{pifont}
\newcommand{\cmark}{\ding{51}}
\newcommand{\xmark}{\ding{55}}

\usepackage{comment}
\usepackage{mathtools}
\usepackage[font=small,skip=6pt]{caption}

\usepackage[onelanguage]{algorithm2e}
\usepackage{xcolor,colortbl}
\usepackage{float}
\usepackage{xcolor}
\usepackage{booktabs}
\definecolor{lightgray}{gray}{0.95}
\definecolor{darkgreen}{rgb}{0.0, 0.2, 0.13}
 \definecolor{darkorchid}{rgb}{0.6, 0.2, 0.7}
\usepackage{dblfloatfix} 
\usepackage{placeins}

\usepackage[symbol]{footmisc}
\def\correspondingauthor{\footnote{Corresponding author.}}

\usepackage[utf8]{inputenc}
\usepackage{framed}
\usepackage{nomencl} 
\makenomenclature
\renewcommand*\nompreamble{\begin{multicols}{1}}
\renewcommand*\nompostamble{\end{multicols}}

\journal{Journal of \LaTeX\ Templates}

\journal{Computer Methods in Applied Mechanics and Engineering}

\usepackage{cleveref} 

\usepackage{hyperref}
\hypersetup{
    colorlinks=true,
   linkcolor=blue,
    citecolor=black,
    pdftitle={phase field cohesive zone modeling},
}

\let\oldhref\href
\renewcommand{\href}[2]{\oldhref{#1}{\hbox{#2}}}

\begin{document}
\begin{frontmatter}




\title{Phase Field Cohesive Zone Modeling for Fatigue \\ Crack Propagation in Quasi-Brittle Materials}


\author{Abedulgader Baktheer$^{1}$}
\author{Emilio Martínez-Pañeda$^{2}$}
\author{Fadi Aldakheel$^{3,}$\correspondingauthor{}}

\address{$^{1}$Institute of Structural Concrete, RWTH Aachen University, 52074 Aachen, Germany }

\address{$^{2}$Department of Engineering Science, University of Oxford, Oxford OX1 3PJ, UK}  

\address{$^{3}$Institute of Mechanics and Computational Mechanics, Leibniz Universität Hannover, 30167 Hannover, Germany \\
Email address: fadi.aldakheel@ibnm.uni-hannover.de}

\begin{abstract} 
The phase field method has gathered significant attention in the past decade due to its versatile applications in engineering contexts, including fatigue crack propagation modeling. Particularly, the phase field cohesive zone method (PF-CZM) has emerged as a promising approach for modeling fracture behavior in quasi-brittle materials, such as concrete.
The present contribution expands the applicability of the PF-CZM to include the modeling of fatigue-induced crack propagation.
This study critically examines the validity of the extended PF-CZM approach by evaluating its performance across various fatigue behaviours, encompassing hysteretic behavior, S-N curves, fatigue creep curves, and the Paris law. The experimental investigations and validation span a diverse spectrum of loading scenarios, encompassing pre- and post-peak cyclic loading, as well as low- and high-cyclic fatigue loading. The validation process incorporates 2D and 3D boundary value problems, considering mode I and mixed-modes fatigue crack propagation.
The results obtained from this study show a wide range of validity, underscoring the remarkable potential of the proposed PF-CZM approach to accurately capture the propagation of fatigue cracks in concrete-like materials. 
Furthermore, the paper outlines recommendations to improve the predictive capabilities of the model concerning key fatigue characteristics.

\end{abstract} 


\end{frontmatter}

\begin{table*}[!t]   
\small

\begin{framed}

\nomenclature[S]{$\Bar{\alpha}(t)
$}{accumulated history variable}
\nomenclature[S]{$\alpha(t)
$}{accumulation variable}
\nomenclature[S]{$t$}{pseudo time}
\nomenclature[S]{$\phi$}{phase field parameter}
\nomenclature[S]{$g(\phi)$}{degradation function}
\nomenclature[S]{$\gamma(\phi, \nabla\phi)$}{crack surface density function}
\nomenclature[S]{$ G_\mathrm{f}$}{fracture energy}
\nomenclature[S]{$f_\mathrm{t} $}{tensile strength}
\nomenclature[S]{$ \ell$}{internal length scale}
\nomenclature[S]{$ a_1, a_2, a_3$}{ parameters controlling the degradation function }
\nomenclature[S]{$p$}{ exponent controlling the degradation function}
\nomenclature[S]{$\Delta a$}{increment of crack length}
\nomenclature[S]{$a$}{crack length}
\nomenclature[S]{$ \Delta N$}{increment of load cycles}
\nomenclature[S]{$ \Delta K_\mathrm{I}$}{amplitude of stress intensity factor}
\nomenclature[S]{$ \Delta K_\mathrm{IC}$}{amplitude of the critical stress intensity factor}
\nomenclature[S]{$E_0$}{Young’s modulus}
\nomenclature[S]{$C, m$}{constants of Paris law}
\nomenclature[S]{$k_0$}{initial slope of the softening regime}

\nomenclature[S]{$N$}{ number of cycles}
\nomenclature[S]{$N^\mathrm{f}$}{ number of cycles to fatigue failure of individual loading range}
\nomenclature[S]{$w$}{crack opening displacement}
\nomenclature[S]{$\psi$}{stored energy density}
\nomenclature[S]{$W$}{work density function}
\nomenclature[S]{$\psi_0$}{initial stored energy density}
\nomenclature[S]{$\mathbf{\sigma}$}{stress tensor}
\nomenclature[S]{$\tilde{\mathbf{\sigma}}$}{effective stress tensor}
\nomenclature[S]{$\mathbf{\varepsilon}$}{strain tensor}
\nomenclature[S]{$\mathbb{C}$}{fourth order elastic stiffness tensor}
\nomenclature[S]{$\mathbb{S}$}{fourth order compliance tensor}
\nomenclature[S]{$\Omega$}{reference
configuration of a cracking solid}
\nomenclature[S]{$\partial \Omega$}{external boundary of $\Omega$}
\nomenclature[S]{$\partial \Omega_\mathrm{t}$}{external boundary of $\Omega$ were tractions are applied}
\nomenclature[S]{$\partial \Omega_\mathrm{u}$}{external boundary of $\Omega$ with prescribed displacements}

\nomenclature[S]{$\mathcal{S}$}{sharp crack}
\nomenclature[S]{$\mathcal{B}$}{localization band}
\nomenclature[S]{$\partial \mathcal{B}$}{external boundary of the localization band}
\nomenclature[S]{$\mathbf{b^*}$}{body forces}
\nomenclature[S]{$\mathbf{t^*}$}{applied tractions}
\nomenclature[S]{$\mathbf{n}$}{outwards unit normal vector}
\nomenclature[S]{$\mathbf{n}_\mathcal{B}$}{outwards unit normal vector of $\mathcal{B}$}
\nomenclature[S]{$\mathbf{n}_\mathcal{S}$}{outwards unit normal vector of $\mathcal{S}$}

\nomenclature[S]{$\mathcal{D}$}{accumulated dissipative part due to fracture}
\nomenclature[S]{$\mathcal{E}$}{stored energy density}

\nomenclature[S]{$\delta$}{variation operator}
\nomenclature[S]{$\nabla$}{gradient operator}
\nomenclature[S]{$\Delta$}{Laplacian operator}
\nomenclature[S]{$Y$}{conjugate energy release rate associated with the crack phase field}
\nomenclature[S]{$\mathcal{Y}$}{thermodynamic driving force}
\nomenclature[S]{$\mathcal{H}$}{history field variable of the driving force}
\nomenclature[S]{$\mathcal{H}_\mathrm{min}$}{minimum value of the fracture driving force}

\nomenclature[S]{$k_\mathrm{f}$}{parameter controlling fatigue damage accumulation}

\nomenclature[S]{$f(\Bar{\alpha})$}{fatigue degradation function}
\nomenclature[S]{$\alpha_\mathrm{T}$}{fatigue damage threshold}

\nomenclature[S]{$\Tilde{\sigma}_1$}{first principal effective stress}
\nomenclature[S]{$\mathbf{u}$}{displacement}
\nomenclature[S]{$S^\mathrm{max}$}{upper level of the loading range}
\nomenclature[S]{$S^\mathrm{min}$}{lower level of the loading range}

\nomenclature[S]{$b, h, L$}{width, height and length of the beam}
\nomenclature[S]{$L_0$}{span of the beam}
\nomenclature[S]{$b_0, h_0$}{width and depth of the notch}
\nomenclature[S]{$F_\mathrm{u}$}{ultimate force}
\nomenclature[S]{$F$}{applied force}

\nomenclature[S]{$\xi$}{coefficient of the geometrical function}
\nomenclature[S]{$c_0$}{scaling parameter}
\nomenclature[S]{$\hat{\alpha}(\phi)$}{geometrical function of the PF-CZM}

\nomenclature[S]{$\mathbf{R}_\mathbf{u}$}{residual vector related to displacement}
\nomenclature[S]{$\mathbf{R}_\phi$}{residual vector related to phase field variable}
\nomenclature[S]{$\mathbf{K}_\mathbf{u}$}{stiffness matrix related to displacement}
\nomenclature[S]{$\mathbf{K}_\phi$}{stiffness matrix related to phase field variable}
\nomenclature[S]{CMOD}{crack mouth opening displacement}
\nomenclature[S]{CMSD}{crack mouth sliding displacement}
\nomenclature[S]{CTOD}{crack tip opening displacement}

\nomenclature[S]{DIC}{digital image correlation}
\nomenclature[S]{SDVs}{solution dependent state variables}
\nomenclature[S]{UMAT}{user material subroutine}
\nomenclature[S]{HETVAL}{heat flux subroutine}

\printnomenclature

\end{framed}
\end{table*}

\section{Introduction}

In various engineering structures, from highway and railroad bridges to airport pavements and offshore structures, the inherent vulnerability to fatigue-induced failure continues to be a critical challenge. This phenomenon, characterized by the progressive propagation of cracks due to cyclic loading, has been studied in detail as it poses a significant concern to the integrity and durability of these essential infrastructures~\citep{bazant1991,  GAN_2021, LI_2020, MIARKA_2022, Baktheer_2022_tip_bearing}.

Over the past few decades, this challenge has motivated researchers to perform extensive experimental studies on fatigue crack propagation in normal and high-strength concrete. These studies include various loading configurations, including mode I, mode II and mixed modes~\citep{baluch1989fatigue, XIAO_2013, GAN_2021, jia_2022, Becks_mode_II, becks_2022_IABSE}. The fatigue crack propagation rate is usually evaluated in the scope of the Paris law~\citep{paris1963critical}, while the fatigue life is often characterised by W\"ohler/S-N curves~\citep{Cornelissen_1984, ZHANG1997}. However, it should be noted that these simple empirical approaches have limited applicability and are insufficient for predicting fatigue behavior at the structural level or under scenarios with complex loading conditions~\citep{nguyen2001, BAKTHEER_2021_4}.

Therefore, a comprehensive analysis of cyclic and fatigue behavior, especially in relation to crack propagation in concrete structures under a wider range of load configurations is of significant importance. This requires the development and application of advanced numerical tools to realistically model the fatigue fracture phenomenon and enable an in-depth analysis~\citep{bazant1993, Becks_2022_monitoring}. This is essential not only for a deep understanding of this intricate process but also as a basis for an economical and reliable design of concrete structures subjected to fatigue loading.

In recent decades, many researchers have adopted the cohesive zone modeling (CZM) approach to model the process of crack propagation in concrete under cyclic loading (e.g.,~\cite{Gylltoft_1984, Hordijk_1991, harper2010fatigue, Zhaodong_2018, Dekker_2019, XI_2022, PARRINELLO_2022}).
However, it is important to recognize that these models tend to represent fatigue crack growth as a one-dimensional mechanism. As a result, their applicability in capturing realistic fatigue-induced degradation is limited for more general loading configurations, such as mode II and mixed mode fracture conditions~\citep{kirane2015, Baktheer_2023}.

To better capture a wide range of loading configurations, several macro-scale continuum models have been developed by many authors~(e.g.,~\cite{BAKTHEER_2019_classification, basaldella2022compressive, baktheer_numerical_2018}). In these, fatigue damage is derived by a strain measure, such as total strain~\citep{mai_2012, titscher2020}, plastic strain~\citep{kindrachuk2015constitutive}, or cumulative strain measure~\citep{desmorat_continuum_2007, kirane2015, BAKTHEER_2021_1, aguilar_2021, Aguilar_framcos_2023}. These models have demonstrated the ability to effectively capture the main aspects of fatigue degradation in quasi-brittle materials.
However, the sensitivity of the numerical results to the finite element size (mesh sensitivity) in simulations involving strain localization due to softening is an issue associated with such local continuum models~\citep{Bazant_2002}. Therefore, these approaches have to be combined with appropriate regularization techniques such as the crack band approach~\cite{bazant1983crack} or non-local approaches of integral- and gradient-type~\citep{kroner_1968, ERINGEN_1972, PEERLINGS_1996}.

The self-regularized phase field method (PFM) has attracted tremendous attention in the past decade, as the crack paths can be automatically determined as a part of the solution~\citep{FRANCFORT_1998, BOURDIN_2000}. 
This approach shows remarkable capabilities in capturing complicated crack phenomena that include aspects such as crack nucleation, branching, and coalescence~\citep{ZHOU_2018, Amirian_2023, Abali_2023}. In particular, the scope extends to various fracture types ranging from brittle to ductile and quasi-brittle behavior~\citep{Miehe_2017, DITTMANN_2018, DAL_2022, Amirian_2022, Zhao_2023}.
Moreover, specific failure surfaces can be easily integrated into the formulation of the variational phase field method, as shown in~\cite{NAVIDTEHRANI_2022, SCHRODER_2022}. 
For these reasons, PFM has been used in various applications, such as multi-physics problems~\citep{CUI_2021,Ai2022,aldakheel2022electro}, micromechanics~\citep{ALDAKHEEL_2020, ULLOA_2022}, porous media~\citep{HEIDER_2021}, and hydrogen embrittlement~\citep{MARTINEZPANEDA_2018,Hageman2023b}.

\begin{figure*}[!t]
\centerline{
\includegraphics[width=1.0\textwidth]{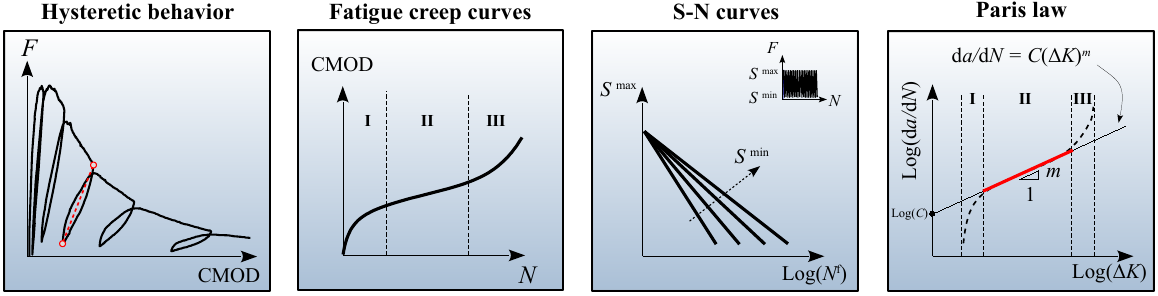}}
\caption{Key characteristics of cyclic and fatigue behavior of concrete based on experimental observations, including the following abbreviations: force ($F$), crack mouth opening displacement (CMOD), upper level of the loading range ($S^\mathrm{max}$), number of cycles ($N$), number of cycles to fatigue failure ($N^\mathrm{f}$),  crack length ($a$), and stress intensity factor ($K$) }
\label{f:fatigue_aspects}
\end{figure*}

Modeling of fatigue crack propagation using PFM has been recently addressed by many authors. In this context, a fatigue degradation function is introduced in~\cite{ALESSI_2018, Carrara_2020} that effectively reduces the fracture toughness of the material due to the externally applied repeated loads. This was achieved by introducing an energy accumulation variable that takes into account the structural loading history. 
A unified framework was introduced in~\cite{selevs2021} to account for fatigue crack propagation in brittle and ductile soils by considering elasto-plastic material behavior. 
A similar approach was recently published in~\cite{GOLAHMAR_2022, KHALIL_2022}, where nonlinear kinematic and isotropic hardening were considered. Further advancements in phase field modeling targeting both brittle and ductile fatigue failure can be explored in~\citep{schreiber_2021, ULLOA_2021, Aldakheel_2022, Yan_2022, KALINA_2023_2}.

Recently, the phase field cohesive zone model (PF-CZM) developed by Wu~\cite{WU_2017} has been successfully used for modeling quasi-brittle materials such as concrete and showed precise capture of crack propagation in concrete under monotonic loading, covering a wide range of stress configurations in 2D and 3D applications~\citep{WU_2021_3D}. This approach was able to capture the concrete size effect and represents a promising approach for modeling crack propagation in quasi-brittle materials, as shown in~\cite{FENG_2018, YU_2023, KOREC_2023, FANG_2023}.

The objective of this work is to introduce a phase-field-based modeling approach for the simulation of fatigue crack propagation in quasi-brittle materials. A primary novelty lies in the extension of the (PF-CZM) approach, which is specifically tailored to the simulation of cyclic and fatigue crack propagation in concrete. Another key aspect of this work is the systematic and comprehensive experimental validation based on experimental programs that include several loading scenarios~\cite{BAKTHEER_2021_4, jia_2022}. This validation includes a critical evaluation of the applicability of the extended (PF-CZM) model over a wide range of loading scenarios. This includes pre-peak and post-peak cyclic loading as well as low- and high-cycle fatigue scenarios. The comprehensive experimental validation demonstrates the robustness and adaptability of the proposed approach.
To the best of our knowledge, such an extension of the PF-CZM and the application of the method to quasi-brittle materials with systematic and comprehensive experimental validation has not been explored in previous literature. 
In this context, we would like to elucidate and highlight the advancements that our work brings to this field making the contribution to the unified modeling of low-cycle fatigue (LCF) and high-cycle fatigue (HCF) in concrete. Fig.~\ref{f:fatigue_aspects} provides a visual summary of the key fatigue aspects considered in the current study.

The paper starts with the formulation of the regularized phase field cohesive zone approach including the extension to fatigue loading in Sect.~\ref{sec:PF-CZM_formulation}. Numerical implementation aspects of the presented approach are briefly described in Sect.~\ref{sec:implementation:aspects}. The numerical studies of fatigue crack propagation covering mode I and mixed modes, including the comparison with experimental results, are shown and described in Sect.~\ref{sec:mode_I}, Sect.~\ref{sec:mixed_mode_I_II}, and Sect.~\ref{sec:mixed_mode_I_III}. Finally, a critical evaluation of the presented approach based on the main fatigue crack propagation aspects in concrete is summarized in Sect.~\ref{sec:evaluation}.

\section{Phase field regularized cohesive zone framework} \label{sec:PF-CZM_formulation}
The phase field formulation for quasi-brittle materials like concrete, which extends the classical phase field model for brittle fracture by including general geometric and energetic functions in line with~\cite{WU_2017, FENG_2018, WU_2021_3D}, is introduced in this section.
We consider a reference configuration of a cracking solid $\Omega \subset \mathbb{R}^n (n \in [1, 2, 3])$, with an external boundary
$\partial \Omega \subset \mathbb{R}^{n-1}$ with outwards unit normal $\mathbf{n}$ (Fig.~\ref{f:PFM}a). 
The localization band denoted by $\mathcal{B} \subseteq \Omega$ encompasses the crack $\mathcal{B}$, and its external boundary is represented by $\partial \mathcal{B}$ with outwards unit normal vector denoted as $\mathbf{n}_\mathcal{B}$.
Assuming small strains $\mathbold{\varepsilon}(\mathbf{u})$, 
in order to quantify both the energy stored in the material and the dissipation, a split of the work density function $W$ into energetic and dissipative parts is defined as:
\begin{equation}\label{eq:PFM_top}
    W(\mathbf{u}, \phi) = \int_{\Omega} \psi [ \mathbold{\varepsilon}(\mathbf{u}), g(\phi) ]~~\mathrm{d}V + \int_{\mathcal{B}} G_\mathrm{f} \; \gamma(\phi, \nabla\phi)~~\mathrm{d}V
\end{equation}
where the first term $ \mathcal{E}(\mathbf{u}, \phi)~=~\int_{\Omega} \psi [ \mathbold{\varepsilon}(\mathbf{u}), g(\phi) ]~\mathrm{d}V$ is the stored energy density and the second term $ \mathcal{D}(\phi, \nabla\phi) = \int_{\mathcal{B}} G_\mathrm{f} \; \gamma(\phi, \nabla\phi)~\mathrm{d}V$ is
the accumulated dissipative part due to fracture~\cite{MIEHE_2016}.
%
The function $ g(\phi)$ represents the degradation function and $\gamma(\phi, \nabla\phi)$ is the crack surface density function. Here, $\phi$ represents the scalar phase field parameter, which varies between 0 (indicating undamaged material) and 1 (representing fully damaged material), and $G_\mathrm{f}$ is the material fracture energy.
\subsection{Geometric crack surface density function:}
The crack surface density function is defined using the crack phase field $\phi$ and its gradient $\nabla\phi$
\begin{equation}\label{eq:gamma_01}
     \gamma(\phi, \nabla\phi)  =  \dfrac{1}{c_0} \left[ \dfrac{1}{\ell} \hat{\alpha}(\phi) + \ell |\nabla\phi|^2 \right] 
\end{equation}
\noindent with the scaling parameter $c_0$ being introduced to ensure that the fracture energy is reproduced~\cite{WU_2017}, defined as
\begin{equation}\label{eq:c_0}
     c_0 = 4 \int_{0}^{1} \sqrt{\hat{\alpha}(\beta)} d\beta
\end{equation}
where $\ell$ is the internal length
scale regularizing the sharp crack and $\hat{\alpha}(\phi)$ is a geometrical function which must satisfy the following 
\begin{equation}
\hat{\alpha}(0) = 0, \qquad \hat{\alpha}(1) = 1
\end{equation}
In the phase field cohesive zone model~\cite{WU_2017} the geometrical function is given by 
\begin{equation}\label{eq:alpha_00}
     \hat{\alpha}(\phi) = \xi \phi + (1-\xi) \phi^2
\end{equation}
The shape of the geometrical function is shown in Fig.~\ref{f:PFM}b for varied values of the coefficient $\xi$.
By considering the suggested value in~\cite{WU_2017} of $\xi = 2$, the geometrical function reads
\begin{equation}\label{eq:alpha_02}
     \hat{\alpha}(\phi) = 2\phi -  \phi^2
\end{equation}
which differs from conventional phase field models, which commonly employ either the linear $\hat{\alpha}(\phi) = \phi$ or the quadratic function $\hat{\alpha}(\phi) = \phi^2$ to describe brittle fracture.

\subsection{Energetic degradation function for cohesive cracks}
\begin{figure*}[!t]
\centerline{
\includegraphics[width=1.0\textwidth]{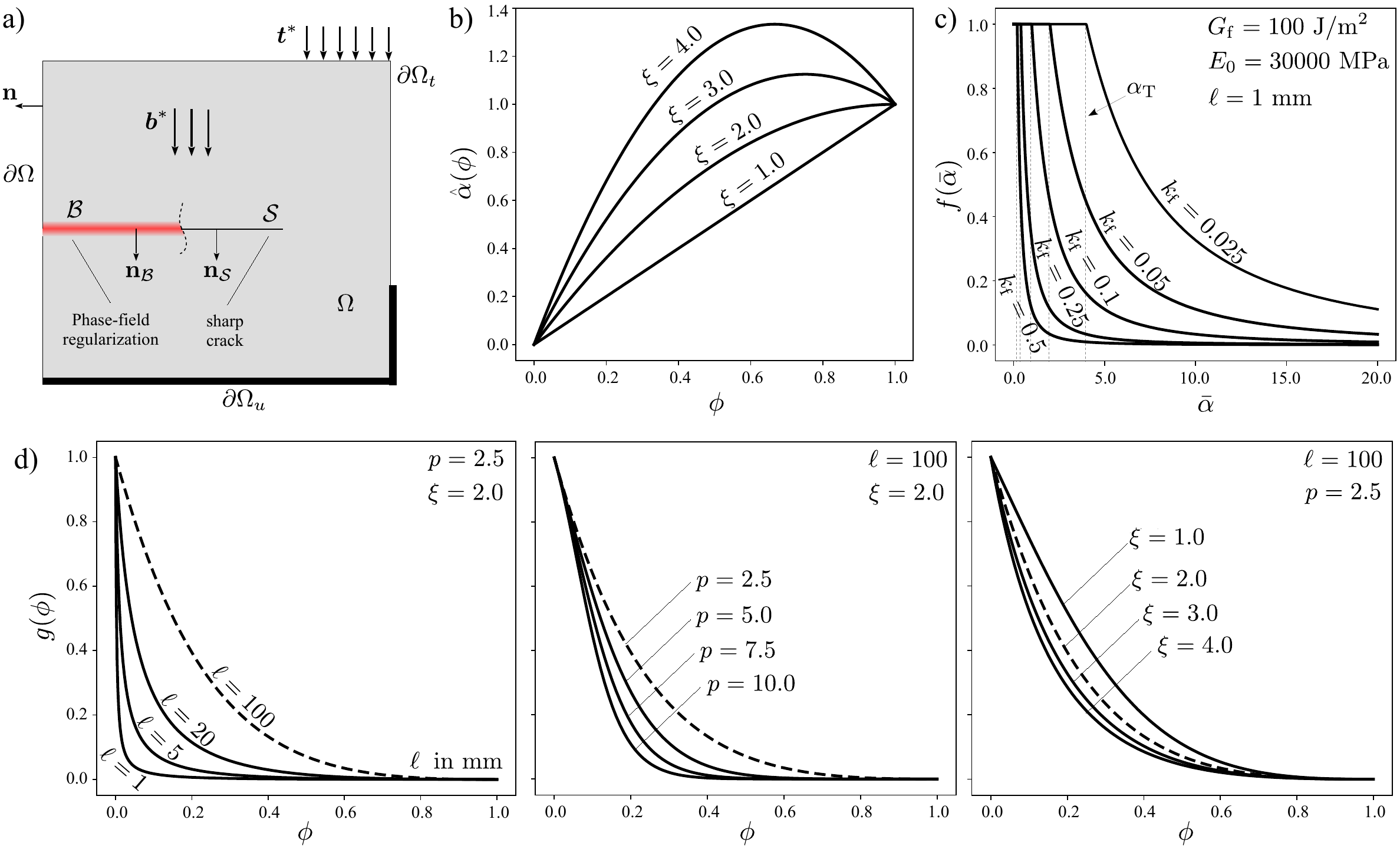}}
\caption{A cohesive zone-based phase field description of fracture and fatigue: a) cracking solid with a sharp crack and phase field regularization;\; b) geometrical function of the phase field cohesive zone model for varied coefficient $\xi$};\; c) fatigue degradation function for several values of the fatigue damage accumulation parameter $k_\mathrm{f}$;\; d) energetic degradation function of the phase field cohesive zone model for varied internal length scale $\ell$, varied exponent $p$ and varied coefficient $\xi$, with fixed material properties of $G_\mathrm{f} = 100 \,\mathrm{J}/\mathrm{m}^2$,\, $E_0 = 30000$ MPa, and $f_\mathrm{t} = 3$ MPa
\label{f:PFM}
\end{figure*}

To describe the degradation of the initial strain energy during the evolution of the crack phase field, a monotonically decreasing energetic function satisfying the following properties shall be defined~\citep{Miehe_2010}.
\begin{equation}\label{eq:g_01}
     g'(\phi) < 0,\quad  g(0) = 1,\quad  g(1) = 0,\quad  g'(1) = 0
\end{equation}
The degradation function is determined using the analysis presented in~\cite{Lorentz_2011}, following a general form that will be specified as
\begin{equation}\label{eq:g_01}
     g(\phi) =  \dfrac{(1-\phi)^p}{(1-\phi)^p + Q(\phi)}
\end{equation}

with 
\begin{equation}\label{eq:Q_01}
     Q(\phi) =  a_1\,\phi + a_1\,a_2\,\phi^2 + a_1\,a_2\,a_3\,\phi^3
\end{equation}
where the parameters $a_1, a_2, a_3, p$ can be determined based on the chosen cohesive law that governs the material softening behavior.
Considering an exponential cohesion law that represents the stress $\sigma$ as a function of the opening displacement $w$, it can be expressed as follows:
\begin{equation}\label{eq:cohesive_law}
     \sigma(w) = f_\mathrm{t} \,\exp(- \dfrac{f_\mathrm{t}}{G_\mathrm{f}} w)
\end{equation}
with an initial slope of the softening regime given as
\begin{equation}\label{eq:k_0}
     k_0 = - \dfrac{{f_\mathrm{t}}^2}{G_\mathrm{f}}
\end{equation}
where $f_\mathrm{t}, G_\mathrm{f}$ are the tensile strength and the fracture energy of the material, respectively.
The material parameters $a_1, a_2, a_3, p$ according to~\cite{WU_2017} can be given as follows
\begin{align}\label{eq:a}
    & a_1 = \dfrac{2 E_0 G_\mathrm{f}}{{f_\mathrm{t}}^2} \cdot \dfrac{\xi}{c_0 \ell}\\ \nonumber \\  \label{eq:b}
    & a_2 = \dfrac{1}{\xi} \left[  (- \dfrac{4 \pi \xi^2 }{c_0} \dfrac{G_\mathrm{f}}{{f_\mathrm{t}}^2} k_0 )^{2/3} 
 + 1 \right] - (p +1)\\ \nonumber \\ \label{eq:c}
    & a_3 = \begin{cases}
            0 & \qquad p > 2 \\  \\ 
            \dfrac{1}{a_2} \left[  \dfrac{1}{\xi} (\dfrac{c_0 w_\mathrm{c}\, f_\mathrm{f}}{2 \pi G_\mathrm{f}})^2 - (1+b) \right] & \qquad p = 2
          \end{cases}
\end{align}
Based on fitting data the parameter $p$ should be set to be $> 2$. A value of 2.5 is selected in all studies in the current work.
Considering the suggested value of $\xi = 2$, we substitute this value into (Eq.~\ref{eq:alpha_00}). Accordingly, through the integration described in (Eq.~\ref{eq:c_0}) we obtain the result of $c_0 = \pi$. 
By substituting these values into (Eq.~\ref{eq:a} - Eq.~\ref{eq:c}), the parameters  $a_1, a_2, a_3$ are obtained as
\begin{equation}
    a_1 = \dfrac{4 E_0 G_\mathrm{f}}{{ \pi \ell f_\mathrm{t}}^2}, \qquad a_2= 2^{5/3} - 3, \qquad a_3= 0
\end{equation}
where $E_0$ is the Young's modulus of the described material.
The shape of the described degradation function is illustrated in Fig.~\ref{f:PFM}d, considering different values of the internal length scale $\ell$, the exponent $p$ and the coefficient $\xi$.

\subsection{Constitutive relations} \label{sec_constitutive_relations}
The stored energy density can be given as
\begin{equation}
     \psi [ \mathbold{\varepsilon}(\mathbf{u}), g(\phi) ] = g(\phi) \, \psi_0(\mathbold{\varepsilon})
\end{equation}
where $\psi_0(\mathbold{\varepsilon})$ is the initial strain energy density given as
\begin{equation} 
     \psi_0(\boldsymbol{\varepsilon}) = \dfrac{1}{2} \mathbold{\varepsilon} : \mathbb{C} : \mathbold{\varepsilon} 
     = \dfrac{1}{2} \mathbold{\tilde{\sigma}} : \mathbb{S} : \mathbold{\tilde{\sigma}} 
\end{equation}
where $\mathbb{C}$ is the elastic stiffness, $\mathbb{S}$ is the compliance, and $\mathbold{\tilde{\sigma}}$ is the effective stress tensor given as
\begin{equation} 
\mathbold{\tilde{\sigma}} = \mathbb{C} : \mathbold{\varepsilon}
\end{equation}

The apparent stress tensor can be then defined as 
\begin{equation}  \label{eq:apparent_stress}
\mathbold{\sigma} = \dfrac{\partial \psi [ \mathbold{\varepsilon}(\mathbf{u}), g(\phi) ]}{\partial \mathbold{\varepsilon}} = g(\phi)\,  \mathbold{\tilde{\sigma}} = g(\phi)\,  \mathbb{C} : \mathbold{\varepsilon}
\end{equation}

The conjugate energy release rate associated with the crack phase field
can be obtained as
\begin{equation} \label{eq:Y_minus}
Y = - \dfrac{ \partial \psi [ \mathbold{\varepsilon}(\mathbf{u}), g(\phi) ]}{\partial \phi} = - g'(\phi)  \mathcal{Y}
\end{equation}
where $\mathcal{Y}$ represents the initial (undamaged) elastic strain energy, which can be written as
\begin{equation} 
\mathcal{Y} = \psi_0(\boldsymbol{\varepsilon}) 
\end{equation}
Note that the minus sign in (Eq.~\ref{eq:Y_minus}) is included to ensure a non-negative value of the conjugate energy release rate, similar to the classical continuum damage models, e.g.,~\cite{desmorat_continuum_2007, ragueneau_thermodynamic-based_2006, CHUDOBA_2022}.
\subsection{Governing equations and damage evolution law}
Taking into account the constitutive relations described above and considering the first variation of work density function (Eq.~\ref{eq:PFM_top}) with respect to the primal kinematic variables $\phi$ and $\mathbf{u}$, we can write
\begin{equation}\label{eq:variation_W}
  \delta  W(\mathbf{u}, \phi) =  \delta \mathcal{E}(\mathbf{u}, \phi)  + \delta  \mathcal{D}(\phi, \nabla\phi) 
\end{equation}
where $\delta$ denotes the variation.
The variation of the stored energy density is given as
\begin{align}  
\delta \mathcal{E}(\mathbf{u}, \phi) 
= \int_{\Omega} \left( \dfrac{\partial \psi}{\partial \mathbold{\varepsilon}} : \nabla^{\mathrm{sym}} \delta \mathbf{u} + \dfrac{\partial \psi}{\partial \phi}  \delta \phi\right)~~\mathrm{d}V
= \int_{\Omega} (\mathbold{\sigma} :  \nabla^{\mathrm{sym}} \delta \mathbf{u} - Y \delta \phi)~~\mathrm{d}V  = \int_{\Omega} \left[\mathbold{\sigma} :  \nabla^{\mathrm{sym}} \delta \mathbf{u} + g'(\phi)  \mathcal{Y} \, \delta \phi \right]~~\mathrm{d}V \label{eq:var_stored}
\end{align}
The variation of energy dissipation $\delta \mathcal{D}$ can be obtained as
\begin{equation}  \label{eq:var_dissipation}
\delta  \mathcal{D}(\phi, \nabla\phi) = \int_{\mathcal{B}} G_\mathrm{f} \; \delta \gamma(\phi, \nabla\phi)~~\mathrm{d}V \geq 0
\end{equation}
where the variation of the crack density function is obtained as
\begin{equation} 
 \delta \gamma(\phi, \nabla\phi) = \dfrac{1}{c_0} \left[ \dfrac{1}{\ell} \alpha'(\phi) \delta \phi + 2 \ell \nabla\phi \cdot \nabla\delta\phi\right]
\end{equation}
The variation of the work density function can be then written as
\begin{equation} \label{eq_variation_W_02}
 \delta  W(\mathbf{u}, \phi) = 
 \int_{\Omega} \left[\mathbold{\sigma} :  \nabla^{\mathrm{sym}} \delta \mathbf{u} + g'(\phi)  \mathcal{Y} \, \delta \phi \right]~~\mathrm{d}V + \int_{\mathcal{B}} G_\mathrm{f} \; \delta \gamma(\phi, \nabla\phi)~~\mathrm{d}V
 \end{equation}
 In line with~\cite{NAVIDTEHRANI_2021_2, WU_2017, MANDAL_2019}, the local force balances can be derived by applying Gauss's divergence theorem and the statement that (Eq.~\ref{eq_variation_W_02}) must hold for any kinematically admissible variations of the virtual quantities. The coupled field equations are therefore given as follows,
 \begin{align} 
& \nabla \cdot \mathbold{\sigma} + \mathbf{b^*} = \mathbf{0}   \qquad \mathrm{in}  \quad \Omega\\ \label{eq:damage_law_01}
& g'(\phi)  \mathcal{Y} +  G_\mathrm{f} \delta_{\phi} \gamma  \geq 0 \qquad \mathrm{in}  \quad \mathcal{B}
\end{align}
It should be noted that (Eq.~\ref{eq:damage_law_01}) represents the damage evolution law.
The variation of the crack surface density function $ \delta_{\phi} \gamma$ is obtained as
\begin{equation} \label{eq:delta_phi_gamma}
\delta_{\phi} \gamma = \partial_{\phi} \gamma - \nabla \cdot \partial_{\nabla \phi} \gamma = \dfrac{1}{c_0} \left[ \dfrac{1}{\ell } \alpha'(\phi) - 2 \ell \Delta \phi \right]
\end{equation} 
with $\Delta \phi$ denoting the Laplacian $\Delta \phi = \nabla \cdot \nabla \phi$.
By substituting (Eq.~\ref{eq:delta_phi_gamma}) into (Eq.~\ref{eq:damage_law_01}) and considering (Eq.~\ref{eq:apparent_stress}), the coupled field equations can be rewritten as follows 
\begin{align}
    &\nabla \cdot \left[g(\phi) \,\mathbf{\Tilde{\sigma}} (\mathbf{u}) \right] + \mathbf{b^*} = \boldsymbol{0}\\[2mm] \label{eq:phase_filed_balance}
    &\dfrac{G_\mathrm{f}}{c_0} \left[ \dfrac{2}{\ell } (1-\phi) - 2 \ell \Delta \phi \right] = -g'(\phi) \mathcal{Y}(\mathbf{u})  \qquad \mathrm{for} \quad  \dot{\phi} > 0
\end{align}
These field equations are supplemented by the following Neumann boundary conditions
\begin{align}
\begin{cases}
     \mathbf{\sigma} \cdot \mathbf{n} = \mathbf{t}^*  \qquad  & \mathrm{on}~~\partial \Omega_\mathrm{t} \\
     \nabla \phi \cdot \mathbf{n}_\mathcal{B} = 0 \qquad  & \mathrm{on}~~\partial\mathcal{B} 
\end{cases}
\end{align}
where $\partial \Omega_\mathrm{t}$ is the part of  $\partial \Omega$ where the tractions are applied.

Two approaches are available for solving the discretized forms of the above field equations. The first one involves a monolithic scheme, where both $\mathbf{u}$ and $\phi$ are simultaneously solved. The second option is a staggered scheme, which employs an alternate minimization strategy. For further details, see e.g.,~\citep{KRISTENSEN_2020, KHALIL_2022}. In the current study, the staggered scheme was used in all numerical simulations described in the following sections.

\subsection{Fracture driving force and irreversibility} \label{sec_driving_force}
As evident from the constitutive relation and the conjugate energy release rate presented above in Sect.~\ref{sec_constitutive_relations}, they demonstrate symmetry in response to tension and compression. Consequently, the material is susceptible to compressive fracture as well. 
To address the above issue, the following ad hoc modified effective driving force $\mathcal{Y}$ is used, as introduced in~\cite{wu_2013}:
\begin{equation} 
\mathcal{Y} =  \dfrac{1}{2 E_0} \langle \Tilde{\sigma}_1 \rangle ^2
\end{equation}
where $\Tilde{\sigma}_1$ is the first principal stress and $\langle \cdot \rangle$ defines the Macaulay brackets (i.e. $\langle x \rangle = \mathrm{max}\,(x, 0)$).
It should be noted that the redefinition of the driving force $\mathcal{Y}$ leads to a hybrid isotropic/anisotropic formulation~\cite{ambati_2015_review}, which results in a loss of variation consistency, as described in~\cite{WU_2020_variationally}. However, this formulation is necessary in the current study to retain the isotropic constitutive relation and at the same time allow the representation of asymmetric tension/compression behavior.

Considering the damage to be an irreversible process ( i.e. $\dot{\phi} \geq 0$), a history field variable $\mathcal{H}$ is introduced ensuring that the following condition is always met
\begin{equation}
    \phi_{t + \Delta t} \geq \phi_{t}
\end{equation}
where $\phi_{t + \Delta t}$ represents the phase field variable in the current time increment, while $\phi_{t}$ is the phase field variable in the previous increment. This history variable must satisfy the Kuhn-Tucker conditions
\begin{equation}
    \mathcal{Y} - \mathcal{H} \leq 0, \qquad  \dot{\mathcal{H}} \geq 0, \qquad  \dot{\mathcal{H}} \cdot (\mathcal{Y} - \mathcal{H}) = 0
\end{equation}
Consequently, the history field for a current time step $t$, over a total time $\tau$, can be written as follows:
\begin{equation}
    \mathcal{H} = \underset{t \in [0, \tau]}{\mathrm{max}} \mathcal{Y} (t)
\end{equation}
It should be noted that, in line with~\cite{Navidtehrani_2021_1}, a minimum value of the fracture driving force is defined as
\begin{equation}
    \mathcal{H}_\mathrm{min} = \dfrac{{f_\mathrm{t}}^2}{2 \, E_0}
\end{equation}

\subsection{Extension to fatigue loading}

To address material degradation under fatigue loading, we follow the concept proposed in~\cite{ALESSI_ULLOA_2023}, where the fatigue degradation function is directly incorporated at the level of Griffith’s energy balance, employing the notion of state-dependent fracture toughness. This approach has also been adopted in the context of functionally graded materials~\cite{CPB2019} and toughness reduction due to embrittlement~\cite{Cui2022}. By adhering to this concept, the accumulated dissipative part due to fracture within the phase field cohesive zone framework is modified as follows:

\begin{equation}
    \mathcal{D}(\phi, \nabla\phi, t) = \int_{0}^{t} f(\Bar{\alpha}(t)) \;G_\mathrm{f} \; \gamma(\phi, \nabla\phi)~~\mathrm{d}t
\end{equation}
where $t$ is the pseudo-time, $\Bar{\alpha}(t)$ is the accumulated history variable. The function $f (\Bar{\alpha}(t))$ defines the fatigue degradation function, which has the following properties
\begin{align}\nonumber
 &f(\Bar{\alpha}(t) \leq\alpha_\mathrm{T})  = 1, \quad f(\Bar{\alpha}(t) > \alpha_\mathrm{T}) \in [0, 1]\\[2mm]
  &f'(\Bar{\alpha}(t)) \leq 0 \qquad \mathrm{in} \qquad 0 \leq f(\Bar{\alpha}(t)) <1
\end{align}
The proposed function in~\cite{Carrara_2020} is used in the present study given as
\begin{equation}
    f(\Bar{\alpha}(t)) = 
    \begin{cases}
1 &\qquad  \mathrm{if} \qquad  \Bar{\alpha}(t) \leq\alpha_\mathrm{T}\\[2mm]
\left( \dfrac{2\, \alpha_\mathrm{T}}{ \Bar{\alpha}(t)  + \alpha_\mathrm{T}}\right)^2
&\qquad  \mathrm{if} \qquad  \Bar{\alpha}(t) > \alpha_\mathrm{T}
    \end{cases}
\end{equation}
where the $\alpha_\mathrm{T}$ is the threshold that controls the initiation of fatigue damage, which is defined as
\begin{equation}
    \alpha_\mathrm{T}= \dfrac{G_\mathrm{f}}{k_\mathrm{f} \; \ell}
\end{equation}
Here, $k_\mathrm{f}$ is the material parameter that controls the rate of fatigue damage accumulation and, consequently, the fatigue life of the material.
It should be noted that other fatigue degradation functions can be used as discussed in~\cite{selevs2021, KALINA_2023}.
The shape of the described fatigue degradation function is shown in Fig.~\ref{f:PFM}c for varied values of parameter $k_\mathrm{f}$.

Considering the fatigue hypothesis introduced in~\cite{marigo1985modelling}, the accumulated history variable $\Bar{\alpha}(t)$ can be defined as
\begin{equation}
    \Bar{\alpha}(t) = 
    \begin{dcases}
        \int_{0}^{t}|\dot{\alpha}| \; \mathrm{d}t
        &\;\;\; \mathrm{if} \;\;\;\;  \alpha \dot{\alpha} \geq 0 \;\;(\mathrm{loading~stage}) \\[2mm]
        0  &\;\;\; \mathrm{if} \;\;\;\;  \alpha \dot{\alpha} < 0 \;\;(\mathrm{unloading~stage})
    \end{dcases}
\end{equation}
Where $\alpha$ is the accumulation variable that has been been chosen in line with~\citep{KHALIL_2022} as $ \alpha(t) = [1 -\phi(t)]^2 \; \psi_0(t)$. 
It should be noted that the choice of this accumulation variable in the context of the presented PF-CM approach here is done arbitrarily to demonstrate the generality of the presented approach; however, other forms of the accumulation variable $\alpha(t)$ can also be chosen, such as $\alpha(t) = g(\phi(t)) \; \psi_0(t)$. Also other specific functions that distinguish between elastic and plastic parts of the strain energy, as discussed in~\cite{ULLOA_2021, selevs2021}, as well as other accumulation strategies, as described in~\citep{GOLAHMAR_2023}, can be employed.

\section{Numerical implementation aspects} \label{sec:implementation:aspects}

\begin{figure}[!t]
\centerline{
\includegraphics[width=\textwidth]{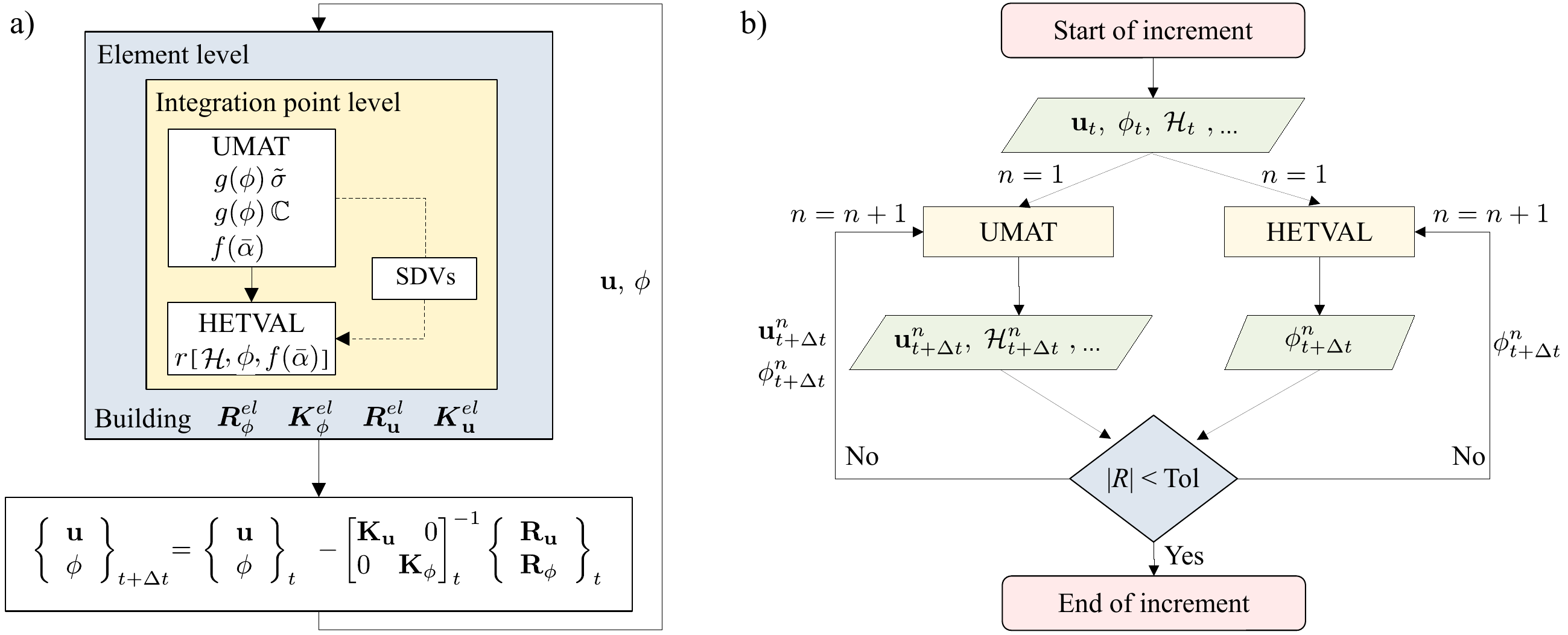}}
\caption{ Implementation concept of the PF-CZM approach: a) flow chart of the subroutine for implementing a coupled deformation-phase field model utilizing the analogy to heat transfer,\; b) flowchart for the solution of the phase field at each integration point for a given increment using a staggered scheme}
\label{fig:flowchart}
\end{figure}

The phase field cohesion zone framework discussed in Sect.~\ref{sec:PF-CZM_formulation} is efficiently implemented within the finite element software ABAQUS using a user materials subroutine (UMAT) at the integration point level and builds on the concepts introduced in~\cite{Navidtehrani_2021_1, NAVIDTEHRANI_2022}. This implementation takes advantage of the equivalence between the heat transfer law and the phase field balance equation (Eq.~\ref{eq:phase_filed_balance}), as visualized in Fig.~\ref{fig:flowchart} and discussed below.

Consider a solid with thermal conductivity $k$, specific heat $c_P$, and density $\rho$. When subjected to a heat source $r$, the temporal evolution of the temperature field $T$ is governed by the following balance law:
\begin{equation}
    k\; \nabla^2 T - \rho c_P \dfrac{\partial T}{\partial t} = - r
\end{equation}
In a steady-state condition, this simplifies to ${\partial T}/{\partial t} = 0$, and the balance law becomes:
\begin{equation}
    k\; \nabla^2 T = - r
\end{equation}

The analogy between this elliptic partial differential equation and the phase field evolution law is evident, with the temperature field serving as the phase field $(T \equiv \phi)$. To highlight this analogy further, the phase field balance equation (Eq.~\ref{eq:phase_filed_balance}) can be reformulated as, considering the historical variable $\mathcal{H}$:
\begin{equation}
   \Delta \phi = \nabla^2 \phi = \dfrac{c_0}{2 \ell G_\mathrm{f}} g'(\phi) \mathcal{H} + \dfrac{1}{\ell^2} (1-\phi) 
\end{equation}

Assuming a unit thermal conductivity $k = 1$, we can define the heat flux due to internal heat generation as:
\begin{equation} \label{eq:heat_flux}
    r = -\dfrac{c_0}{2 \ell G_\mathrm{f}} g'(\phi) \mathcal{H} - \dfrac{1}{\ell^2} (1-\phi) 
\end{equation}

For the computation of the Jacobian matrix, it is necessary to define the rate of change of heat flux $r$ with respect to temperature
\begin{equation} \label{eq:heat_flux_derv}
    \dfrac{\partial r}{\partial \phi} = -\dfrac{c_0}{2 \ell G_\mathrm{f}} g''(\phi) \mathcal{H} + \dfrac{1}{\ell^2}. 
\end{equation}

It should be noted that, to account for fatigue degradation, the fracture energy $G_\mathrm{f}$ in (Eq.\ref{eq:heat_flux}) and (Eq.\ref{eq:heat_flux_derv}) is replaced by the fatigue-degraded quantity $f(\Bar{\alpha}(t)) ;G_\mathrm{f}$.

The application of the heat transfer analogy in ABAQUS can be easily accomplished by utilizing the user material (UMAT) and heat flux (HETVAL) subroutines. This approach allows for implementation at the integration point level, taking advantage of built-in displacement-temperature elements like the ABAQUS CPE4T type, suitable for 4-node bilinear quadrilateral elements.
At integration point level in ABAQUS, the subroutines receive strain and phase field values (temperature) that are interpolated from nodal solutions for a specific element. The user material subroutine (UMAT) is first called in each integration point loop and enables the calculation of the material stiffness tensor $\mathbb{C}$ and the stress tensor $\mathbold{\sigma}$ from the strain tensor. The current value of the phase field is then used to take into account the damage degradation of these tensors. 
The accumulated history variable and the fatigue degradation function are then calculated.
The driving force is stored in solution-dependent state variables (SDVs), enforcing irreversibility, as well as the fatigue degradation function. The updated SDV values are transferred to the heat flux subroutine (HETVAL), which is used to transfer the current value of the history field $\mathcal{H}$ and fatigue degradation function $f(\Bar{\alpha})$. In the HETVAL subroutine, the internal heat flux $r$ (Eq.~\ref{eq:heat_flux}) and its derivative with respect to temperature (phase field) (Eq.~\ref{eq:heat_flux_derv}) are defined, considering the fatigue degraded fracture energy. This process is repeated for each integration point so that ABAQUS can construct element stiffness matrices and residuals, resulting in the assembly of the global system of equations as shown in Fig.~\ref{fig:flowchart}a.

Within the utilized staggered solution scheme, an alternative minimization approach, namely of the single-pass type, is used for the sequential solving of the displacement subsystem, $\mathbf{K}_\mathbf{u} \, \mathbf{u} = \mathbf{R}_\mathbf{u}$, and the phase field subsystem, $\mathbf{K}_\phi \, \phi = \mathbf{R}_\phi$. The residual and the stiffness matrix for the phase field subsystem are constructed considering the historical field from the previous increment $\mathcal{H}_t$. In other words, the history field remains frozen during the entire iterative process, as illustrated in Fig.~\ref{fig:flowchart}b. 
It is important to note that the fatigue degradation function and the accumulated history variables are handled in a similar manner to the history field variable, which is omitted in Figs.~\ref{fig:flowchart}b for simplicity.
The utilized subroutines, complete with usage instructions and a demonstrative example, have been made accessible to the reader and can be found in~\cite{Baktheer_2024_ABAQUS_code}.
For more comprehensive insights into the ABAQUS implementation, please refer to~\cite{NAVIDTEHRANI_2021_2, Navidtehrani_2021_1}.
Additionally, it is worth noting that an alternative approach for implementing the phase field model in ABAQUS involves utilizing the user element subroutine (UEL), a method demonstrated by several authors such as~\cite{MSEKH_2015, PILLAI_2018, SELES_2019, KRISTENSEN_2020}.

\section{Mode I crack propagation}
\label{sec:mode_I}

To evaluate the validity of the phase field cohesive zone approach presented in Sect.~\ref{sec:PF-CZM_formulation}, this section focuses on numerical simulations that examine the propagation of mode I cracks within concrete. These simulations cover scenarios involving monotonic, cyclic, and fatigue loading.

\subsection{Experimental program}

We follow the test program by Baktheer and Becks~\cite{BAKTHEER_2021_4} on notched concrete beams as it encompassed a wide range of loading conditions~\cite{Baktheer_2021_3}.

\paragraph{Geometry, material properties, and measurement techniques}

\begin{figure}[!t]
\centerline{
\includegraphics[width=\textwidth]{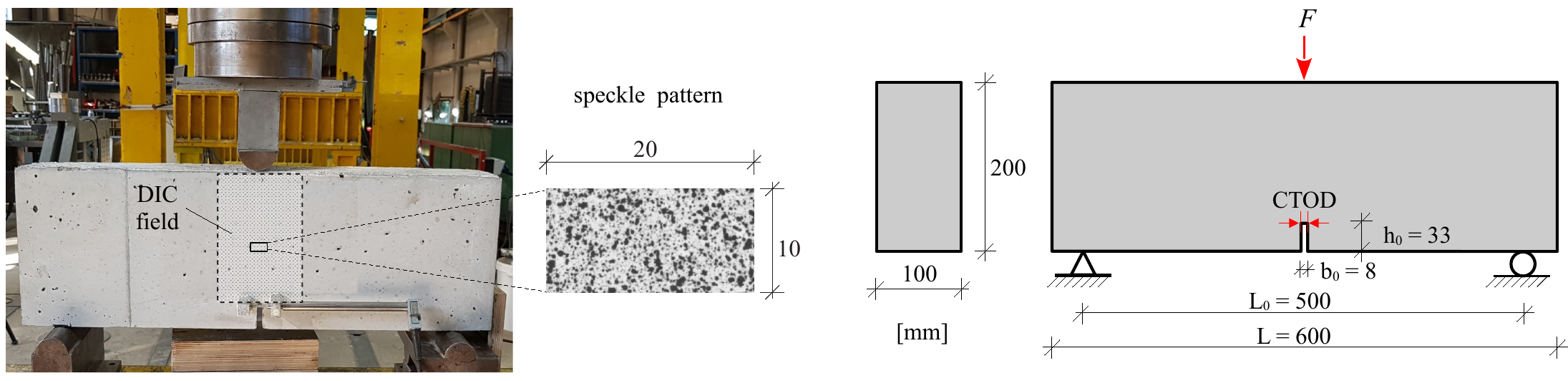}}
\caption{Test setup and geometry of the notched three-point bending tests of concrete (mode I crack propagation)}
\label{fig:mode_I_test_setup}
\end{figure}

The three-point bending tests employed notched beam specimens with a cross-section height ($h$) of 200~mm and a beam width ($b$) of 100~mm. The dimensions of the beam length and span were proportionally adjusted based on the beam height, as shown in Fig.~\ref{fig:mode_I_test_setup}. The notch depth ($h_0$) was established at $h/6$, while the notch width ($b_0$) was set to 8~mm.
The concrete used in this study falls under the concrete strength class C60, according to~\cite{MC10}. 
During testing, the crack tip opening displacement (CTOD) was measured using linear differential transformers (LVDT). In addition to the LVDT measurements, a 3D digital image correlation (DIC) measurement system was used in selected tests to measure crack propagation during cyclic and fatigue loading.

\begin{table*}[!t]
\renewcommand*{\arraystretch}{1.35}
	\small
  \centering
  \setlength\extrarowheight{-5pt}
    \caption{Description of the loading scenarios used in the experimental program}
  \begingroup\setlength{\fboxsep}{0pt}
  \colorbox{lightgray}{%
  \begin{tabular}{  m{2.6cm}  m{4.6cm}  m{6.5cm}   m{1.9cm}  }
    \toprule
    \textbf{Loading scenario} & \textbf{Description} & \textbf{Purpose} & \textbf{Figure} \\ \toprule
    LS1
    & 
    Monotonic loading   
    &
    \raggedright{Studying the monotonic behavior}
    &
    \includegraphics[scale=0.35]{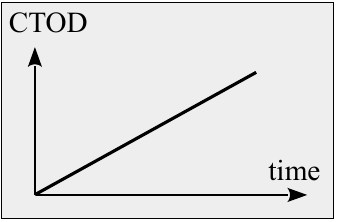}
    \\ \hline
    LS2
    & 
    \raggedright{Post-peak cyclic loading}
    &
    \raggedright{Providing detailed information of
    unloading and reloading behavior in the post-peak regime }
    &
    \includegraphics[scale=0.35]{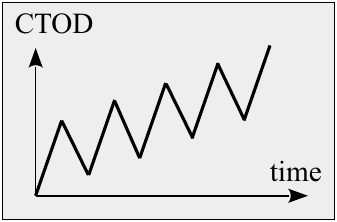}
    \\ \hline
    LS3
    & 
    \raggedright{Pre-peak cyclic with step-wise increased loading }
    &
    \raggedright{Providing detailed information of
    unloading and reloading behavior in the pre-peak regime}
    &
    \includegraphics[scale=0.35]{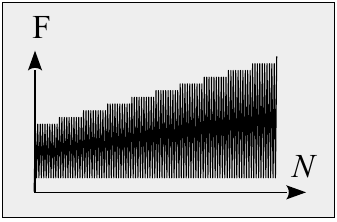}
    \\ \hline
    LS4
    & 
    \raggedright{Fatigue loading with constant amplitudes}
    &
    \raggedright{Studying the low- and high-cycle fatigue behavior under constant amplitudes}
    &
    \includegraphics[scale=0.35]{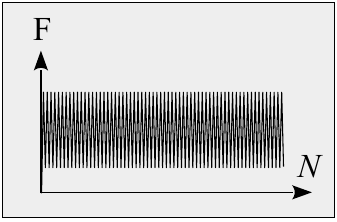}
    \\ \toprule 
  \end{tabular}%
}\endgroup
\\
{\footnotesize It should be noted that the first and second loading scenarios are displacement control tests, controlled by crack tip opening displacement (CTOD) over time, while the third and fourth loading scenarios are force (F) control tests, displayed through applied loading cycles}
\label{tbl:loading_scenarios}
\end{table*}

\paragraph{Loading scenarios}

The test campaign involved a series of loading scenarios to investigate the cyclic flexural and fatigue properties of concrete. These scenarios are outlined in Table~\ref{tbl:loading_scenarios} and serve as the basis for a systematic calibration and validation protocol for numerical models predicting fatigue crack propagation in concrete.
The first loading scenario (LS1) presents a typical monotonically increasing loading pattern controlled by the crack tip opening displacement (CTOD).
The second loading scenario (LS2) describes a cyclically increasing pattern for the upper and lower loading levels, as outlined in Table~\ref{tbl:loading_scenarios}, which is controlled by the CTOD. This scenario includes five unloading cycles and provides detailed insight into the post-peak loading and unloading response.
The third scenario (LS3) involves cyclic loading controlled by the applied load, encompassing a maximum of 100 cycles. The upper load level starts at $S^\mathrm{max} = 0.50$ and incrementally increases by $\Delta S = 0.05$, spanning 10 cycles within each load level as indicated in Table~\ref{tbl:loading_scenarios}. The lower load level remained constant at $S^\mathrm{min} = 0.10$.
This loading scenario allows a detailed evaluation of the loading and unloading response at the fatigue relevant subcritical load levels before the peak in accelerated form.
The fourth loading scenario (LS4) incorporates the conventional fatigue loading scenario with constant amplitudes. Two different load ranges were investigated, characterized by upper levels of $S^\mathrm{max} = 0.70$ and $S^\mathrm{max} = 0.85$. The lower load levels were accordingly set at $S^\mathrm{min} = 0.05$.

\subsection{Numerical Simulations}
\label{sec:mode_I_numerical}

The three-point bending tests in the experimental program, including the four loading scenarios, were simulated using the Phase field cohesive zone (PF-CZM) approach.
The material parameters for the simulations included Young's modulus ($E_0$), Poisson's ratio ($\nu$), tensile strength ($f_t$), fracture energy ($G_f$), and the fatigue parameter $k_f$. 
The tensile strength and the fracture energy have been calibrated to fit the monotonic response (LS1), while the parameter $k_\mathrm{f}$ has been calibrated based on the fatigue tests (LS4).

Extensive verifications have shown that the length scale parameter $\ell$ has a negligible effect on global responses and crack patterns as long as it remains sufficiently small, e.g.,~\citep{WU_2017, MANDAL_2019, HAI_2021}. Mesh-objective results were obtained by ensuring that the characteristic finite element length was five times smaller than the phase field length scale $\ell$~\cite{PTRSA2021}.
A value of $\ell = 2.5$~mm was deliberately chosen based on the recommendations in~\cite{WU_2017, WU_2021_3D}, which is appropriate for effective simulation of crack propagation in concrete.
A plane stress state is assumed for the 2D model of the notched beam. To ensure accurate representation, a fine mesh is used especially near the notch. Triangular elements with a size of 0.5~mm are used in this mesh configuration. The entire beam is discretized with a total of 8741~elements.
The boundary conditions in the 2D model were applied directly to the nodes, with fixed horizontal and vertical displacements for the node at the bottom left corner of the beam and a fixed vertical displacement for the node at the bottom right corner. The load was applied in the middle of the beam cross-section from the top through a steel plate with a width of 15~cm and a height of 10~cm and elastic material behavior.

\begin{figure}[!t]
\centerline{
\includegraphics[width=\textwidth]{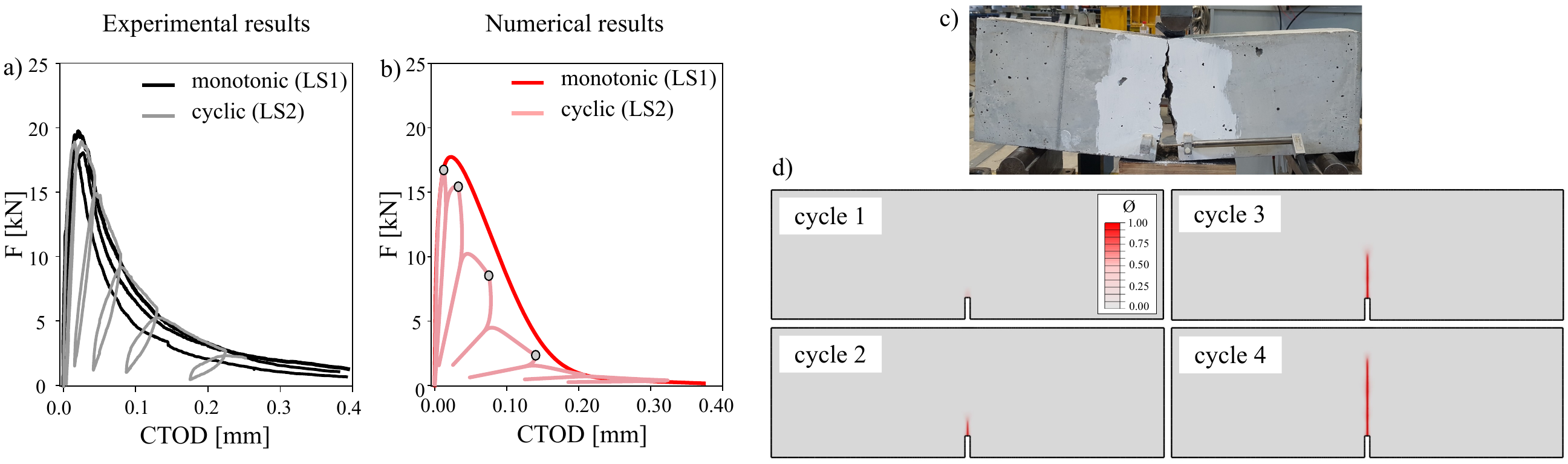}}
\caption{Monotonic and post-peak cyclic response of concrete under mode I crack propagation: a) experimental response (including three monotonic tests (LS1) and one cyclic test (LS2))};\; b) simulated response;\; c) experimental crack path at failure under post-peak cyclic loading (LS2);\; d) crack phase filed propagation under post-peak cyclic loading (LS2) plotted at the first four load cycles
\label{fig:mode_I_results_LS1_LS2}
\end{figure}

\paragraph{Monotonic behavior}
The material parameters were calibrated to obtain a reasonable fit of the monotonic response, which has the following values:  
Young's modulus $E_0 = 30000$~MPa, Poisson's ratio $\nu=0.2$, tensile strength $f_\mathrm{t} = 4.8$~MPa and fracture energy $G_\mathrm{f} = 0.03$~N/mm. The numerically obtained response is shown in Fig.~\ref{fig:mode_I_results_LS1_LS2}b, which represents the load-CTOD curve. The experimental results are shown in Fig.~\ref{fig:mode_I_results_LS1_LS2}a, plotted for three performed tests.
It is worth noting that a better fit can also be achieved by modifying the degradation function to follow the Cornelissen softening law~\citep{cornelissen_1986}, which provides more flexibility in controlling the softening branch, as discussed in~\cite{WU_2017}. 
Nevertheless, the calibrated response can be considered acceptable and appropriate for a quasi-brittle material such as concrete, since the focus of the paper is on the cyclic and fatigue response using the phase field cohesion zone approach.

\paragraph{Post-peak cyclic behavior}

To evaluate the validity of the PF-CZM for reproducing the post-peak cyclic response, a comparison of the experimental response and simulation is shown in Figs.~\ref{fig:mode_I_results_LS1_LS2}a and b, respectively. 
It is essential to acknowledge that the PF-CZM is a pure continuum damage-based model, and therefore, it cannot capture the experimentally observed irreversible crack opening.
It should be noted that irreversible crack opening refers to the unclosed crack due to the rough surface of the developed crack, as experimentally observed once the force reaches zero level.
To address this limitation in cyclic behavior, an extension of the model is required, possibly incorporating plasticity, as seen in general formulations of the phase field method for ductile fracture~\citep{ULLOA_2021, selevs2021, ALDAKHEEL_2020_TAFM, ALDAKHEEL_2021_global_local}. 
Another aspect of the cyclic behavior of concrete is the hysteretic loops. The current formulation of the approach presented considers a pure damage model with the Marigo fatigue hypothesis~\citep{marigo1985modelling} (see Sect.~\ref{sec:PF-CZM_formulation}), which allows for accumulation of damage during the loading phase. It is also assumed that the hysteretic loops cannot be captured in the current form of the model, as shown in Fig.~\ref{fig:mode_I_results_LS1_LS2}b.

It should be noted that such limitations of the current formulations of the PF-CZM can be improved by incorporating plasticity with kinematic hardening and considering a different fatigue hypothesis that links the dissipative processes of damage and plasticity, as shown in~\cite{desmorat_continuum_2007, baktheermicroplane, BAKTHEER_2021_1, Aguilar_2022}.
That enables the accumulation of the plastic strains under constant amplitude cyclic loading, which is known as the ratcheting mechanism. This mechanism has been successfully introduced within the phase field approach in the unified framework for low and high cycle fatigue proposed in \cite{ULLOA_2021}.
The experimentally derived crack propagation is shown in Fig.~\ref{fig:mode_I_results_LS1_LS2}c for a test performed under the LS2 loading scenario. Besides, the numerical representation of the phase field crack propagation during cyclic loading is shown in Fig.~\ref{fig:mode_I_results_LS1_LS2}d, specifically for the first four loading cycles.

\begin{figure*}[!t]
\centerline{
\includegraphics[width=1.0\textwidth]{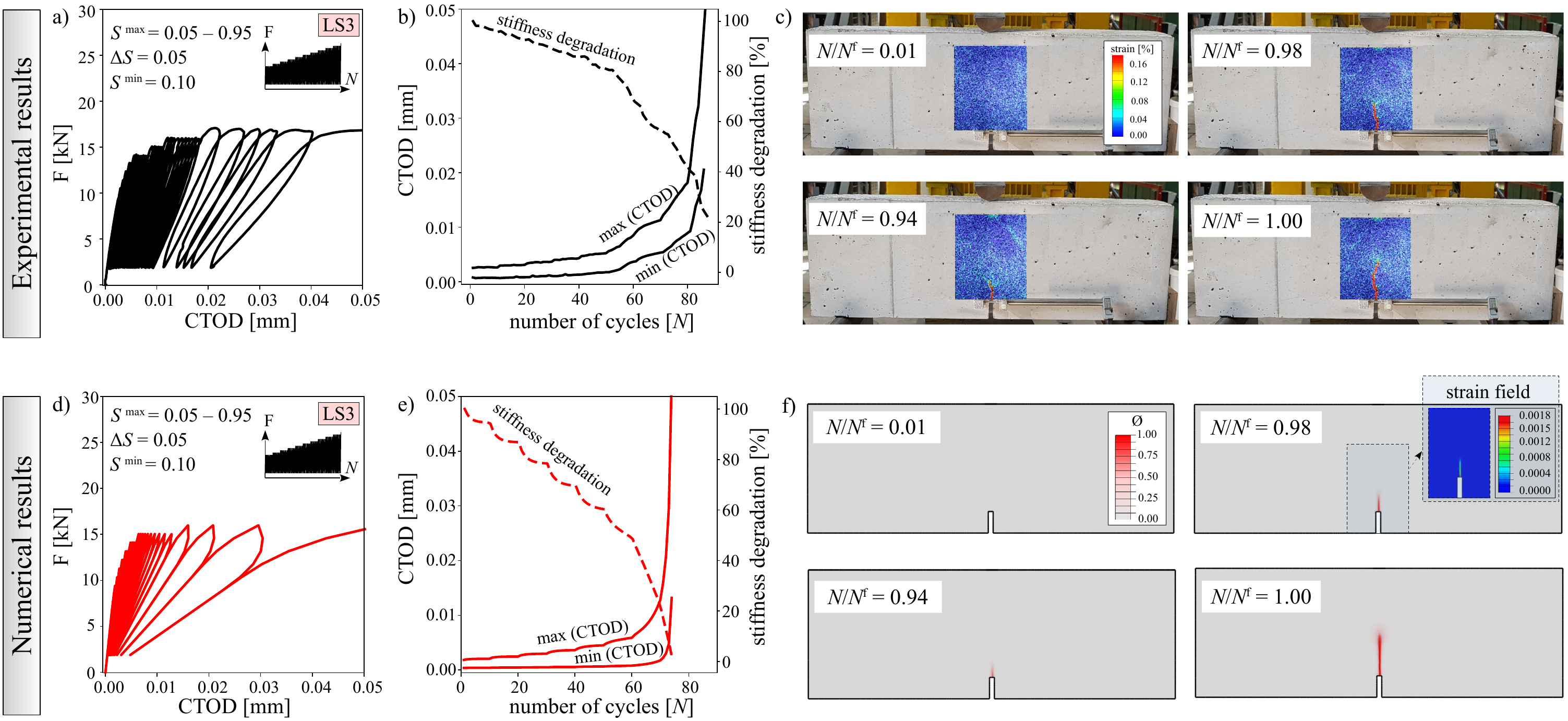}}
\caption{Mode I crack propagation in concrete under pre-peak step-wise cyclic loading: a,d) load-CTOD response obtained experimentally and from simulation, respectively;\; b, e) growth of CTOD at the upper and lower load levels with the corresponding stiffness degradation obtained experimentally and from simulation, respectively;\; c) experimental crack propagation at four stages during the cyclic response measured with DIC;\; f) corresponding numerical phase field crack propagation including the numerically obtained strain field for the third stage}
\label{fig:mode_I_results_LS3}
\end{figure*}

\paragraph{Pre-peak cyclic behavior} 
The response under pre-peak cyclic loading (LS3) simulated with the presented PF-CZM approach compared to the experimental results is shown in Fig.~\ref{fig:mode_I_results_LS3}. Ten load increments per cycle were used in this simulation and all other fatigue simulations. 
Since the applied loads in cyclic and fatigue experiments are relatively slow and controlled, all cyclic and fatigue simulations are assumed to be quasi-static, where the inertial effects are neglected, in line with several fatigue simulations presented in the literature~\cite{Carrara_2020, alliche2004, kirane2015}.
Furthermore, the presented version of the model is considered to be time-independent, where creep and loading rate effects are not considered. Such an extension of the model will be possible by incorporating viscoelasticity and viscoplasticity behavior into the presented approach, however, this is beyond the scope of this study.

The calibrated parameters from the monotonic response were used in this simulation (i.e. $E_0, \nu, f_\mathrm{t}, G_\mathrm{f}$). The fatigue parameter $k_\mathrm{f}$ was calibrated based on the loading scenario LS4, as described in the next sections.
The predicted response shown in the second row of Fig.~\ref{fig:mode_I_results_LS3} failed after 74 load cycles, while the test failed after 86 load cycles, which can be considered acceptable.
The shape of the load-CTOD response shown in Fig.~\ref{fig:mode_I_results_LS3}d agrees reasonably well with the experimental response shown in Fig.~\ref{fig:mode_I_results_LS3}a.
The corresponding fatigue creep curves at upper and lower load levels plotted in solid lines are shown in Fig.~\ref{fig:mode_I_results_LS3}b and f for the experimental and numerical results, respectively. 
The resulting stiffness decrease is shown in dashed lines (Figs.~\ref{fig:mode_I_results_LS3}b, e).

To demonstrate the accuracy of the PF-CZM approach in capturing crack propagation, the phase field crack diagrams are shown in Fig.~\ref{fig:mode_I_results_LS3}f at four stages during the fatigue life.
It can be seen that the onset of crack growth corresponds to 94\,\% of the fatigue life. Then, the crack propagates to the middle of the beam height at failure.
Similar observations have been recorded in the experiment monitored with digital image correlation~\citep{BAKTHEER_2021_4, SEEMAB_2023}. These evaluated results are shown in Fig.~\ref{fig:mode_I_results_LS3}c.
For comparing the strain field in the numerical simulation with the DIC-recorded strain field, Fig.~\ref{fig:mode_I_results_LS3}f displays a numerically obtained strain field plot in the longitudinal direction at the stage that corresponds to 98\,\% of the fatigue life, and shows quantitatively similar results to the experimentally recorded values.

It is important to highlight that, except for the disparity in crack opening values at the lower load levels $S^\mathrm{min}$, the PF-CZM model demonstrates remarkable consistency with the experimental results in predicting pre-peak cyclic behavior. This robust agreement underscores the significant potential of the PF-CZM approach in accurately depicting fatigue crack propagation under a wide range of loading conditions.

\begin{figure*}[!t]
\centerline{
\includegraphics[width=1.0\textwidth]{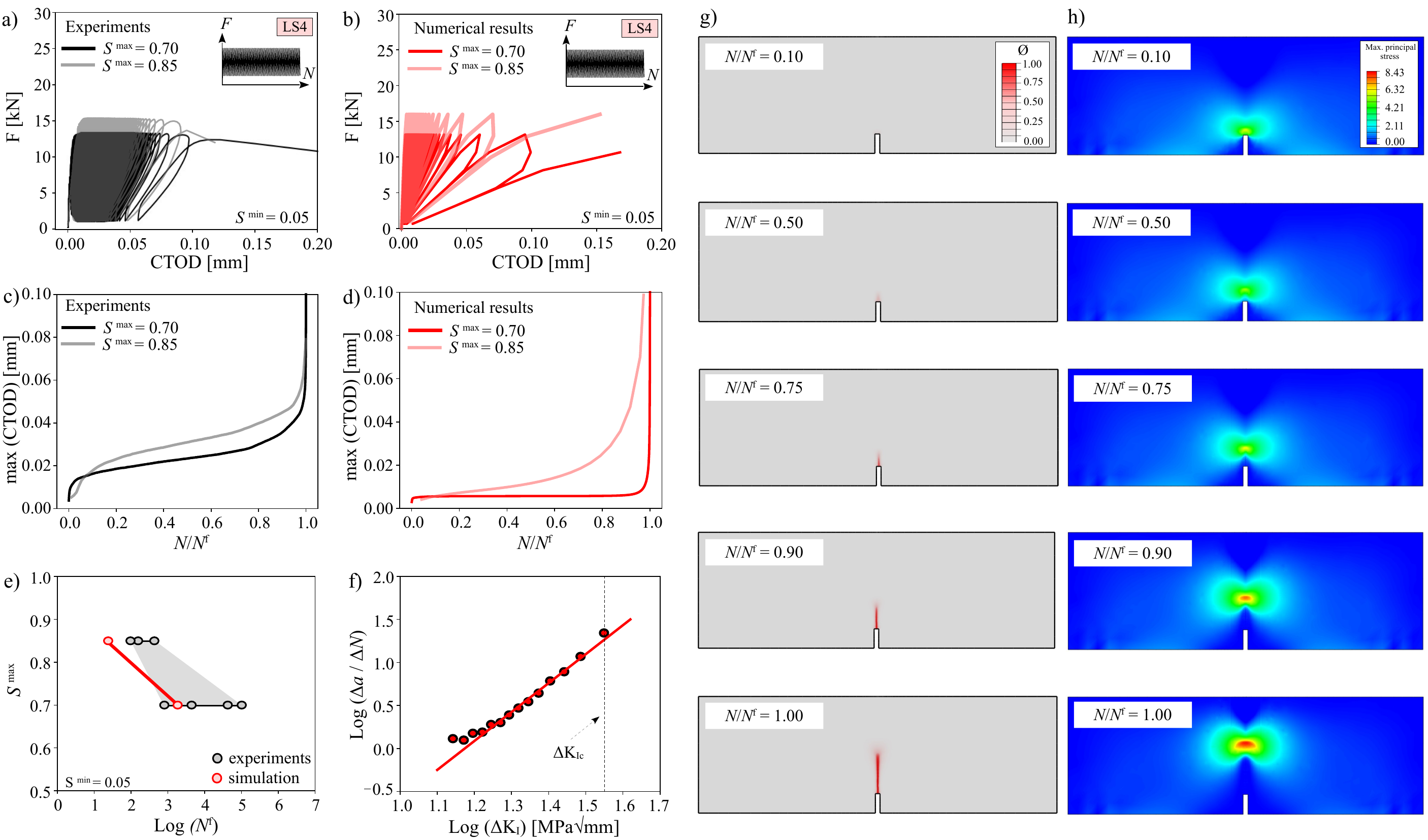}}
\caption{Fatigue response of concrete under mode I crack propagation: a, b) load-CTOD response obtained experimentally and from simulation, respectively;\; c, d)  growth of CTOD at the upper load level obtained experimentally and from
simulation, respectively;\; e) S-N curve (experimental results and numerical results);\; f) numerical results of the crack growth rate (Paris law);\; g) phase field crack propagation during fatigue life;\; h) contours of maximum principal stress during fatigue life}
\label{fig:mode_I_results_LS4}
\end{figure*}

\paragraph{Fatigue behavior}

Concrete behavior under typical fatigue loading with constant amplitudes (LS4) has been simulated, encompassing two distinct loading ranges. The higher load range has an upper level of $S^\mathrm{max} = 0.85$, while the lower load range is characterized by an $S^\mathrm{max} = 0.70$. Notably, both of these load ranges share an identical $S^\mathrm{min} = 0.05$. 
The comparison between the numerical and experimental results is presented in Fig.~\ref{fig:mode_I_results_LS4}.
Within this comparison, we examine the numerical and experimental load–CTOD responses in Figs.~\ref{fig:mode_I_results_LS4}a and b, respectively. Additionally, the corresponding fatigue creep curves are illustrated in Figs.~\ref{fig:mode_I_results_LS4}c and d.
It can be seen that the numerical results exhibit less crack opening displacement throughout the fatigue life in comparison to the experimental results, especially for the lower load range with $S^\mathrm{max} =0.70$.
This discrepancy can be attributed to the inherent limitations of purely damage-based models, which are unable to capture irreversible crack opening displacement.
Similar to the previously discussed issues with hysteretic loops in this section, addressing the accumulation of plastic deformations during cyclic behavior (i.e. the ratcheting mechanism) in combination with hardening effects will enhance the accurate representation of fatigue creep curves.

The calibrated S-N curve, featured in Fig.~\ref{fig:mode_I_results_LS4}e, is presented with the parameter $k_\mathrm{f}$ set to 0.01. It is important to emphasize that further refinement in calibration is attainable by reducing the parameter $k_\mathrm{f}$, resulting in extended fatigue life, particularly for the load range characterized by $S^\mathrm{max} = 0.70$. 
These results demonstrate the applicability of the presented approach to capture the high-cycle fatigue behavior in quasi-brittle materials. This has been achieved by combining the PF-CZM with a appropriate damage accumulation hypothesis. However, it should be noted that in the case of very high-cycle fatigue up to several million load cycles, the integration of the presented approach into a multiscale approach might be necessary, where the connection to the lower scales can be established in order to detect and trigger (nano-micro) defects and their influence on the macroscale fatigue behavior.

\begin{figure}[!t]
\centerline{
\includegraphics[width=\textwidth]{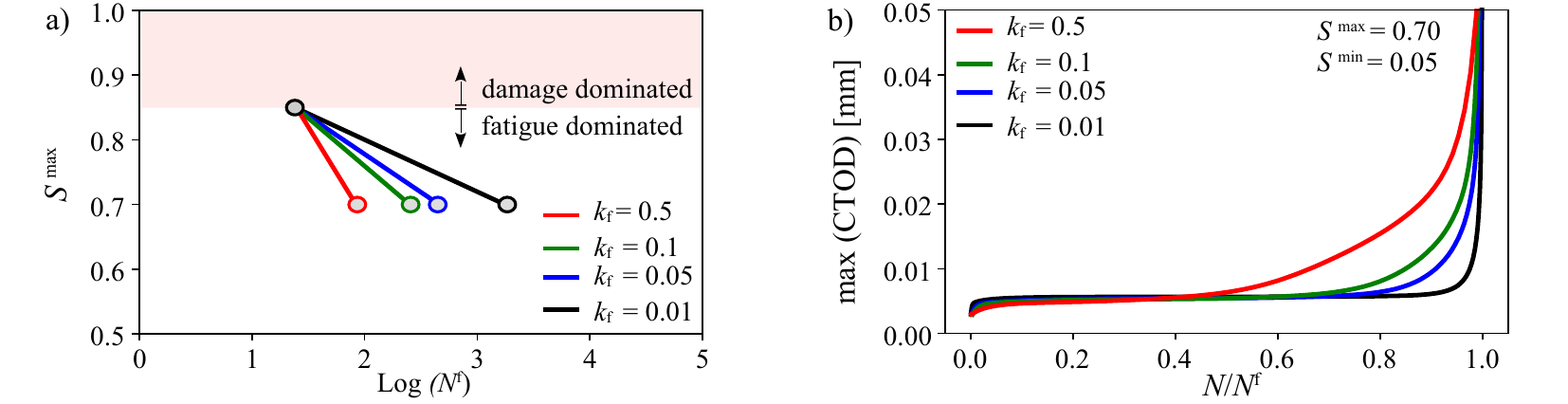}}
\caption{Parametric study for varied value of the fatigue damage accumulation parameter $k_\mathrm{f}$: a) S-N curves;\; b) CTOD growth during fatigue life captured at the upper load level}
\label{fig:mode_I_PS}
\end{figure}

It is worth noting that the adjustments of the fatigue accumulation parameter $k_\mathrm{f}$ have no bearing on the obtained fatigue life within the upper loading ranges, where $S^\mathrm{max} \geq 0.85$.
This issue is vividly illustrated through a parametric study, as showcased in Fig.~\ref{fig:mode_I_PS}. Eight simulations were conducted, spanning both load ranges at $S^\mathrm{max} = 0.70$ and $S^\mathrm{max} = 0.85$, incorporating four distinct values of $k_\mathrm{f}$.
The results reveal that the fatigue life at $S^\mathrm{max} = 0.70$ is subject to precise control through the parameter $k_\mathrm{f}$, as depicted in Fig.~\ref{fig:mode_I_PS}a, along with the corresponding fatigue curves, as evidenced in Fig.~\ref{fig:mode_I_PS}b.

However, an intriguing observation emerges when examining the four simulations at $S^\mathrm{max} = 0.85$: their fatigue lifetimes remained identical. This phenomenon underscores the fact that the fatigue life at this particular load level remains impervious to alterations in the fatigue degradation function. This anomaly can be attributed to the inherent characteristics of the fatigue degradation mechanism at higher load levels, where the dominance of damage processes prevents complete development. This phenomenon, as discussed in~\cite{Carrara_2020}, results from the proximity of the nucleation and unstable crack growth phases, inhibiting the establishment of stable crack growth.
Fortunately, this limitation can be effectively circumvented by adopting alternative fatigue hypotheses, as explored in~\cite{BAKTHEER_2021_1, CHUDOBA_2022, baktheer_2022_monotonic}, allowing for precise control over the S-N curve.

To visualize the simulated fatigue crack propagation, the phase field crack evolution from the simulation at $S^\mathrm{max} = 0.85$ and $k_\mathrm{f} = 0.01$ is shown in Fig.~\ref{fig:mode_I_results_LS4}g. Additionally, the corresponding maximum principal stress contours are provided in Fig.~\ref{fig:mode_I_results_LS4}h.
For a comprehensive evaluation of the ability of the presented PF-CZM approach to capture the Paris law~\citep{paris1963critical}, the crack growth rate from the same simulation with $S^\mathrm{max} = 0.85$ is meticulously analyzed, as illustrated in Fig.~\ref{fig:mode_I_results_LS4}f.
Here, $\Delta K_\mathrm{I}$ represents the amplitude of the stress intensity factor, which is calculated following the methods explained in~\cite{bazant1991, bazant1993}.
The measurement and evaluation of the crack length increment at each loading cycle was carried out manually in the presented example shown in Fig.~\ref{fig:mode_I_results_LS4}f. Automated approaches such as those presented in~\cite{KRISTENSEN_2020} should be used in future work.
As evident from Fig.~\ref{fig:mode_I_results_LS4}f, the plotted data points fit remarkably well along a single line, aligning with the principles of the Paris law. This alignment is consistent with numerous experimental observations on concrete specimens, as corroborated in studies such as~\citep{bazant1991, baluch1989fatigue, CHEN_2022}.
It should be noted that within 
the experimental program used to study mode I fatigue propagation in Fig.~\ref{fig:mode_I_results_LS4} no specific results on the behavior of the Paris law were provided. Therefore, only the numerical results are shown in Fig.~\ref{fig:mode_I_results_LS4}f.

\section{Mixed mode I+II crack propagation}
\label{sec:mixed_mode_I_II}
To assess the validity of the phase field cohesive zone approach presented in Sect.~\ref{sec:PF-CZM_formulation} under more complex loading configurations, numerical simulations focused on the propagation of mixed mode I+II cracks in concrete are presented in this section. These simulations involve monotonic and fatigue loading.

\subsection{Experimental program}
\begin{figure}[!t]
\centerline{
\includegraphics[width=\textwidth]{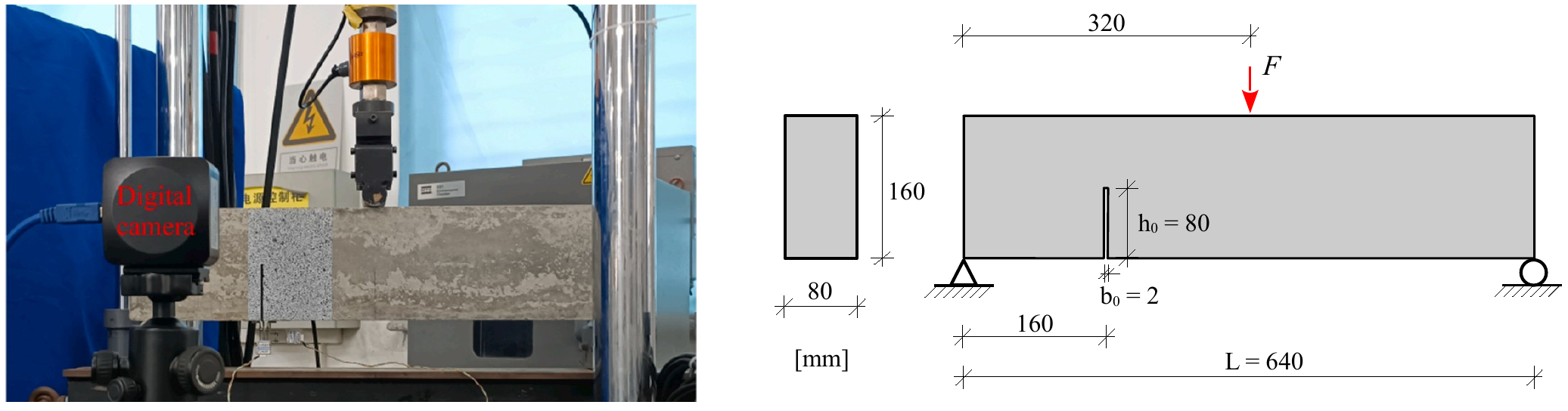}}
\caption{Test setup and geometry of the notched three-point bending tests of concrete (mixed mode I+II crack propagation)~\cite{JIA_2022_EFM}}
\label{fig:mixed_mode_test_setup}
\end{figure}

The test campaign presented in~\cite{jia_2022, JIA_2022_EFM} has been considered for the assessment of the PF-CZM approach, as it covers a reasonable range of parameters under fatigue loading.
The test specimens were notched beams subjected to three-point bending loading. These beams featured a cross-section height ($h$) of 160~mm, a beam width ($b$) of 80~mm, and a span length ($L_0$) of 640~mm. The notch, with a depth ($h_0$) of 80~mm and a width ($b$) of 2~mm, was strategically positioned at a distance of 160~mm from the mid-span cross-section, as illustrated in Fig.~\ref{fig:mixed_mode_test_setup}. The concrete used in these beams had a compressive strength of 44.24~MPa.
During testing, two critical parameters were meticulously monitored. The crack mouth opening displacement (CMOD) and the crack mouth sliding displacement (CMSD) were measured using linear variable differential transformers (LVDT). Additionally, a digital image correlation system was employed to record and analyze the evolving crack patterns.
Within this test program, the analysis focused exclusively on monotonic loading (LS1) and fatigue loading (LS4) scenarios. 
The fatigue tests encompassed five distinct loading ranges, with $S^\mathrm{max}$ values of 0.90, 0.85, 0.80, 0.75, and 0.70, while maintaining a consistent lower load level ($S^\mathrm{min}$) of 0.05.

\begin{figure}[!b]
\centerline{
\includegraphics[width=0.95\textwidth]{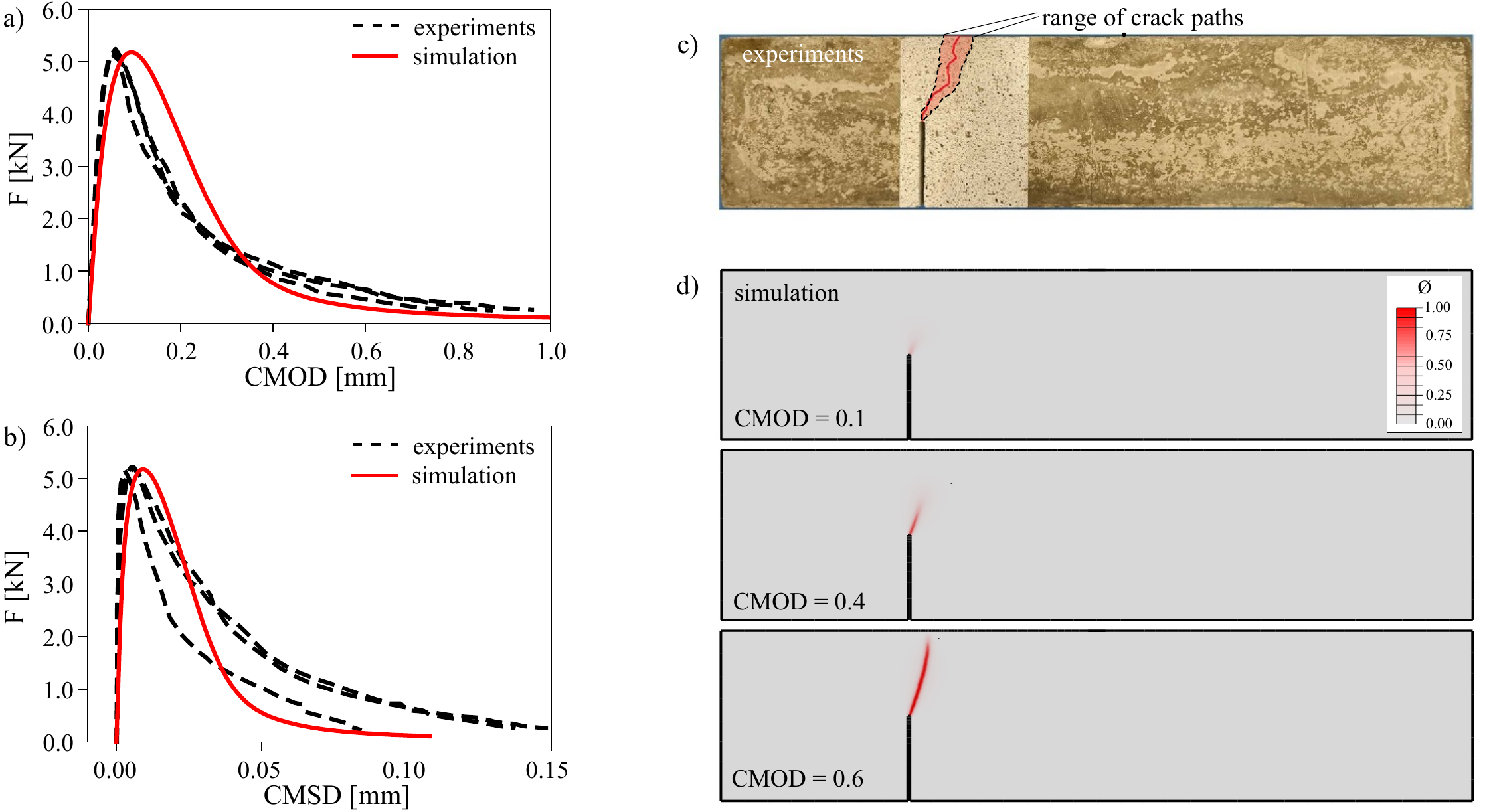}}
\caption{Monotonic response of concrete under mixed model I+II crack propagation (experimental results vs. simulation); a) load-CMOD curve\; b) load-CMSD curve;\; c) experimentally observed crack path;\; d) numerically obtained phase field crack path plotted at three load stages }
\label{fig:mixed_mode_results_LS1}
\end{figure}

\begin{figure*}[!t]
\centerline{
\includegraphics[width=1.0\textwidth]{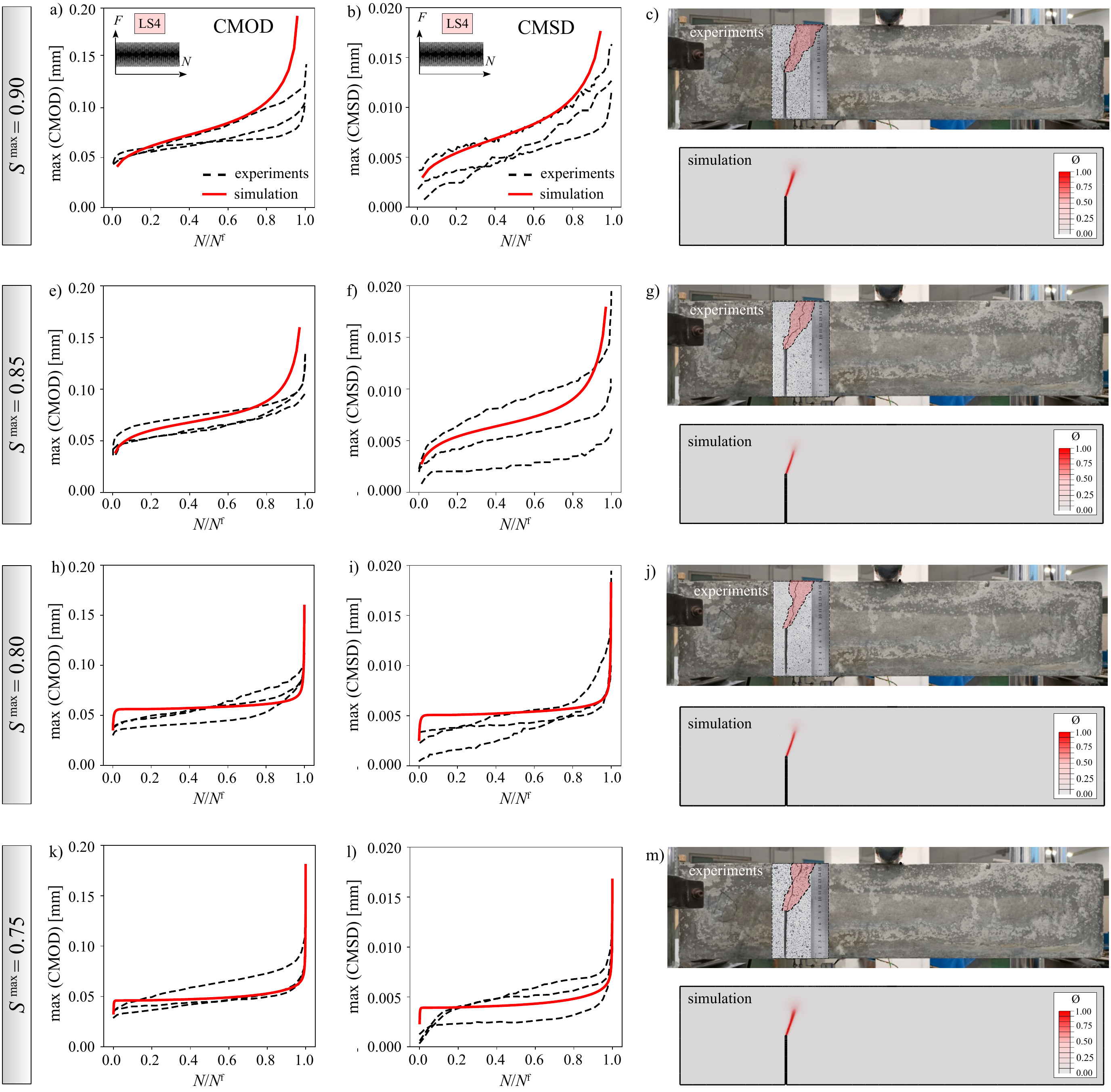}}
\caption{Mixed mode I+II crack propagation under fatigue loading (simulations vs. experimental results presented in~\cite{JIA_2022_EFM}): a, e, h, k) CMOD growth during fatigue life for varied $S^\mathrm{max}$;\; b, f, i, l) CMSD growth during fatigue life for varied $S^\mathrm{max}$;\; c, g, j, m) crack propagation path at fatigue failure}
\label{fig:mixed_mode_results_LS4}
\end{figure*}

\begin{figure*}[!t]
\centerline{
\includegraphics[width=1.0\textwidth]{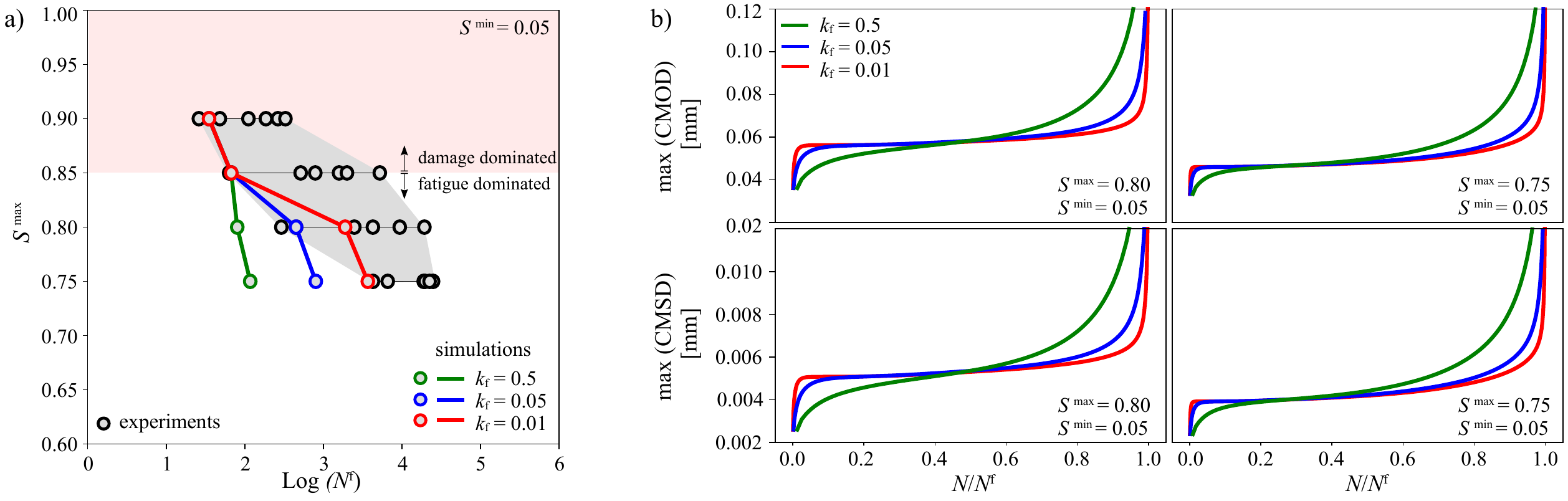}}
\caption{a) S-N curves (experimental results and numerical results for varied $k_\mathrm{f}$);\; b) CMOD and CMSD growth during fatigue life for varied values of $k_\mathrm{f}$ and $S^\mathrm{max}$}
\label{fig:mixed_mode_LS4_SN_PS}
\end{figure*}

\subsection{Numerical modeling}
\label{sec:mixed_mode_I_II_modeling}

\paragraph{Monotonic behavior}
To ensure an accurate representation of the material response under monotonic loading conditions, the model was calibrated using data from monotonic tests. The objective was to achieve a close fit with the experimental load-CMOD and load-CMSD curves, as illustrated in Figs.~\ref{fig:mixed_mode_results_LS1}a and b, respectively.
The calibrated material parameters included a tensile strength ($f_\mathrm{t}$) of 2.8~MPa and a fracture energy ($G_\mathrm{f}$) of 0.09~N/mm. The Young's modulus and Poisson's ratio were set to 30000~MPa and 0.2, respectively. 
The length scale parameter $\ell$ was chosen equal to 2.5~mm consistent with the simulations of mode I crack propagation described in Sect.~\ref{sec:mode_I_numerical}. 
The 2D model assumed a plane stress condition.
Similarly to the simulations in Sect.~\ref{sec:mode_I_numerical}, the boundary conditions were applied directly to the nodes--fixed horizontal and vertical displacements for the node at the bottom left and a fixed vertical displacement for the node at the bottom right. The load is applied in the center of the beam from the top with a 15 cm x 10 cm steel plate.

To evaluate the performance of the PF-CZM approach, the numerically predicted crack propagation path is visualized at three key stages during monotonic loading in Fig.~\ref{fig:mixed_mode_results_LS1}d. 
This visualization is accompanied by the range of experimentally observed crack paths, as shown in Fig.~\ref{fig:mixed_mode_results_LS1}c. 
It is noteworthy that the crack path predicted by the PF-CZM approach closely aligns with the experimentally observed crack path, indicating the capability of the model to reproduce the concrete behavior under monotonic loading conditions.

\paragraph{Fatigue behavior}
To assess the validity of the presented PF-CZM approach for a wide range of loading parameters, simulations of the concrete under mixed mode I+II fatigue crack propagation with four loading ranges have been considered in the study shown in Fig.~\ref{fig:mixed_mode_results_LS4}.
The loading ranges with $S^\mathrm{max}$ of 0.75, 0.80, 0.85 and 0.90 are considered. The lower load level $S^\mathrm{min}$ has a value of 0.05 for the four loading ranges. 
The left part of Fig.~\ref{fig:mixed_mode_results_LS4} shows the obtained CMOD and CMSD growth during fatigue life compared to the experimental results presented in~\cite{jia_2022, JIA_2022_EFM}.

Remarkably, the numerical responses obtained for all four simulated loading ranges exhibit a very good agreement with the experimentally recorded data regarding crack opening and sliding, specifically CMOD and CMSD. This robust agreement reaffirms the capability of the PF-CZM approach to effectively model fatigue crack propagation under mixed mode I+II loading, demonstrating its versatility over a wide spectrum of loading parameters.
The shape of the crack opening growth during fatigue life, as reproduced by numerical simulations, exhibits a characteristic "S" shape. It demonstrates a rapid increase in the initial and final phases, while a linear increase is observed in the middle phase, consistent with previous findings~\citep{jia_2022}. Furthermore, the PF-CZM approach correctly captures the increase in slope observed in the CMOD and CMSD curves during the middle phase with rising values of $S^\mathrm{max}$.
It should be noted that the hysteretic behavior of the simulations shown in Fig.~\ref{fig:mixed_mode_results_LS4} was very similar to the mode I simulations shown in Fig.~\ref{fig:mode_I_results_LS4}b, so these curves are not shown in this section.

In the right part of Fig.~\ref{fig:mixed_mode_results_LS4}, a comparison is presented between the numerically obtained crack paths and the experimentally monitored range of crack paths at the point of fatigue failure. This comparison showcases the capability of PF-CZM approach in predicting mixed mode I+II fatigue crack propagation.
It is worth noting that the phase field of the crack propagation is depicted during the last load cycle, just before a significant deformation increase induced by fatigue failure. As a result, the crack propagates to the middle of the concrete ligament during this final load cycle. In contrast, the visual representation of experimental crack propagation pertains to specimens that have already undergone complete failure.

For the calibration of the fatigue accumulation parameter $k_\mathrm{f}$, the experimentally determined S-N ranges depicted in Fig.~\ref{fig:mixed_mode_LS4_SN_PS}a are considered. 
Three different values of the parameter $k_\mathrm{f}$ are evaluated. A reasonable fit of the fatigue life can be achieved with a value of $k_\mathrm{f}= 0.01$, as demonstrated in Fig.~\ref{fig:mixed_mode_LS4_SN_PS}a.
It is essential to note that loading ranges exceeding $S^\mathrm{max} = 0.85$ are primarily governed by damage, and the fatigue life can not be regulated through the fatigue degradation function, as discussed in Sect.\ref{sec:mode_I_numerical}. 

To achieve full flexibility to match experimentally observed S-N curves, particularly in the high-cycle fatigue regime, the recently introduced fatigue accumulation strategy by Golahmar et al.~\citep{GOLAHMAR_2023} can be utilized and integrated into the PF-CZM framework. Furthermore, this strategy enables the definition of an endurance limit for the fatigue response and provides a flexible way of incorporating the effect of the mean load on the S-N curves.

Fig.~\ref{fig:mixed_mode_LS4_SN_PS}b presents the corresponding CMOD and CMSD growth during the fatigue life for the loading ranges with $S^\mathrm{max} = 0.80$ and $S^\mathrm{max} = 0.75$. Notably, with an increased value of the parameter $k_\mathrm{f}$, the slope of the middle phase of the CMOD and CMSD curves experiences an increase.

\section{Mixed mode I+III  3D crack propagation}
\label{sec:mixed_mode_I_III}

\begin{figure}[!b]
\centerline{
\includegraphics[width=0.7\textwidth]{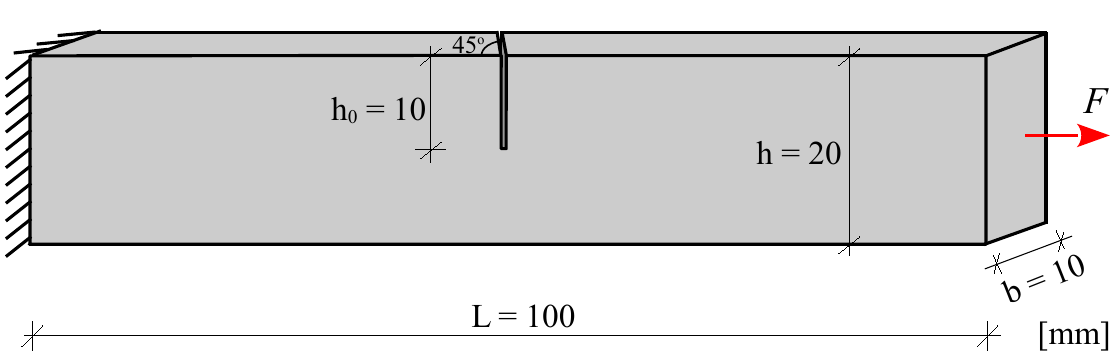}}
\caption{Geometry of the notched 3D beam under uniaxial tension (mixed mode I+III crack propagation)}
\label{fig:mixed_mode_I_III_test_setup}
\end{figure}

In order to assess the capabilities of the presented PF-CZM approach in simulating fatigue crack propagation within 3D applications characterized by complex crack topologies, an example involving a three-dimensional beam with a slanted notch under uniaxial tension is presented in this section.
The geometry of the beam with an inclined notch rotated by $45^{\circ}$ is shown in Fig.~\ref{fig:mixed_mode_I_III_test_setup}. It is important to emphasize that the dimensions of the beam for this illustrative example were deliberately chosen to be as small as possible. The choice is primarily motivated by the goal of minimizing computational costs since there is no experimental data on concrete beams that have been subjected to such tests.

\begin{figure*}[!t]
\centerline{
\includegraphics[width=1.0\textwidth]{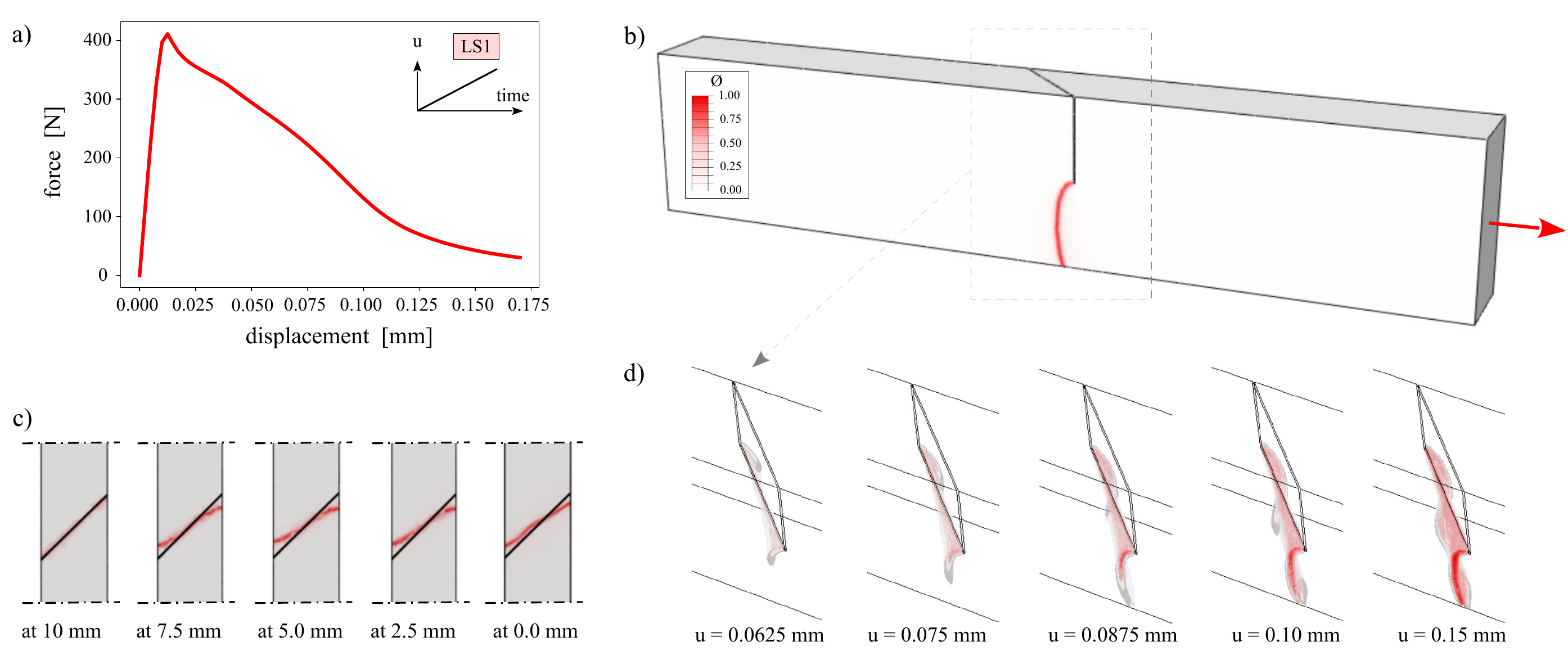}}
\caption{3D numerical example under monotonic loading: a) load-displacement curve;\; b) corresponding phase field crack propagation at failure;\; c) damage profiles at various heights below the notch ;\; d)  3D view of the crack surface at five load stages}
\label{fig:3D_notched_beam_monotonic}
\end{figure*}

\begin{figure*}[!t]
\centerline{
\includegraphics[width=1.0\textwidth]{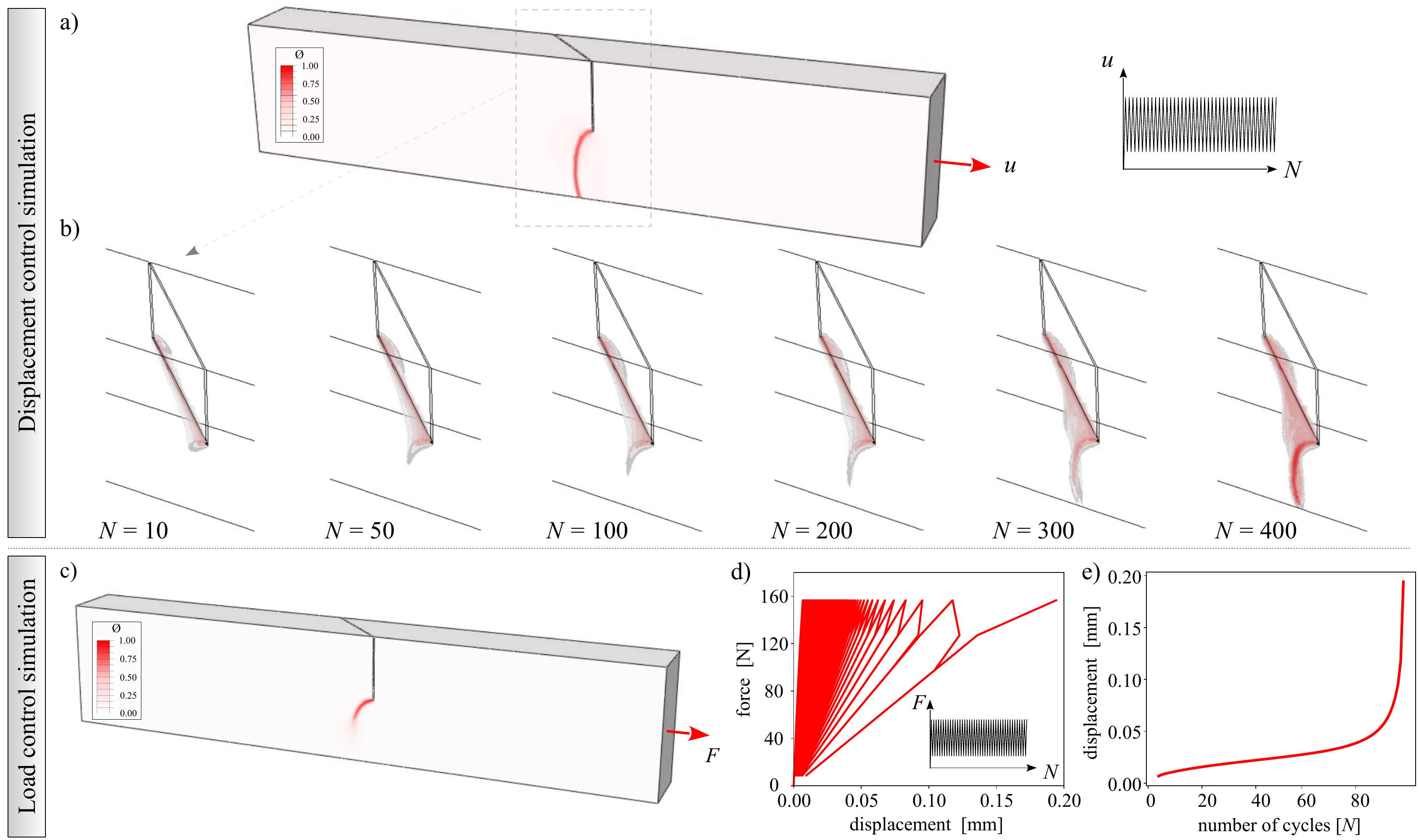}}
\caption{3D numerical example under fatigue loading: a) phase field crack propagation under displacement control fatigue loading at failure;\; b) 3D view of the crack surface at several loading cycles;\; c) phase field crack propagation under load control fatigue loading plotted at failure;\; d) load-displacement response obtained numerically under fatigue loading} ;\; e) corresponding fatigue creep curve 
\label{fig:3D_notched_beam_fatigue}
\end{figure*}

Both monotonic and fatigue simulations have been conducted, as illustrated in Fig.~\ref{fig:3D_notched_beam_monotonic} and Fig.~\ref{fig:3D_notched_beam_fatigue}. 
The phase field length scale for these simulations was set to 0.5 mm, while the material parameters were selected as follows: $E_0 = 30000$ MPa, $\nu = 0.2$, $f_\mathrm{t} = 3.0$ MPa, $G_\mathrm{f} = 0.1$ N/mm, and $k_\mathrm{f} = 0.01$. Tetrahedral elements were employed to discretize the 3D beam, with a refined mesh near the notch having an element size of 0.05 mm. The total number of elements used in the 3D model was 85230.
The boundary conditions with fixed displacements in all three directions were applied directly to the left side of the 3D notched beam, as shown in Fig.~\ref{fig:mixed_mode_I_III_test_setup}, while the load was applied to the entire right side.

\paragraph{Monotonic behavior}
Fig.~\ref{fig:3D_notched_beam_monotonic}a presents the obtained monotonic load-displacement response of the 3D beam. 
In Fig.~\ref{fig:3D_notched_beam_monotonic}b, the 3D phase field crack at the final loading stage can be observed.
During the simulation, the crack initiated from the pre-existing notch, and as the applied load increased, a noticeable rotation of the crack surface was observed. These results effectively capture the anticipated behavior of crack propagation, where the crack gradually re-positions itself to align perpendicular to the direction of maximum principal stress.

Fig.~\ref{fig:3D_notched_beam_monotonic}c offers a top view of the phase field crack at various heights below the notch. The crack initiates from the $45^{\circ}$ slanted notch and then twists over the height of the beam until it returns to its initial position at the $45^{\circ}$ slanted notch.
In Fig.~\ref{fig:3D_notched_beam_monotonic}d, a three-dimensional perspective of the crack surface and its progressive propagation throughout the monotonic loading can be explored. This visualization presents the crack evolution across five distinct load stages, providing a comprehensive depiction of its development.

\paragraph{Fatigue behavior}

To simulate the 3D crack propagation topology under fatigue loading, two simulations were conducted, encompassing both displacement control and load control scenarios, as summarized in Fig.~\ref{fig:3D_notched_beam_fatigue}. A similar virtual experiment was recently conducted by Kristensen et al.~\cite{Kristensen2023} to test accelerated phase field fatigue techniques.
In the displacement control simulation, the 3D phase field crack propagation at fatigue failure is depicted in Fig.~\ref{fig:3D_notched_beam_fatigue}a. 
The loading scenario maintains a constant amplitude with an upper displacement level of 0.095~mm and a lower level of 0.005~mm. For further insight into the 3D crack propagation during the fatigue life, a perspective view at different load cycles is provided in Fig.~\ref{fig:3D_notched_beam_fatigue}b.

Conversely, the fatigue response under load control is demonstrated in the load control simulation. Here, the constant amplitude has an upper load level of 156.6~kN and a lower load level of 8.7~kN. 
The cyclic load-displacement behavior obtained is shown in Fig.~\ref{fig:3D_notched_beam_fatigue}d. The displacement is measured at the point in the upper right corner of the beam.
This response is represented by the displacement growth extracted at the upper load level and is shown in Fig.~\ref{fig:3D_notched_beam_fatigue}e. The corresponding phase field crack at the last loading cycle is presented in Fig.~\ref{fig:3D_notched_beam_fatigue}c.
In contrast to the previous 2D cases described in Sect.~\ref{sec:mode_I_numerical} and Sect.~\ref{sec:mixed_mode_I_II_modeling}, this specific example could not be simulated with a simplified 2D idealization due to the complexity of the crack behavior and the lack of out-of-plane symmetry. The results achieved with the PF-CZM approach clearly demonstrate its capability to capture the intricate crack topology under both monotonic and fatigue loading conditions.

\section{
Critical evaluation and suggestions for further model improvement}
\label{sec:evaluation}

The numerical examples of fatigue crack propagation in concrete presented in the previous sections allow a critical evaluation of the PF-CZM approach presented with respect to the following key fatigue characteristics (as summarized in Table~\ref{t:PFM_CZ_evaluation}):

\paragraph{S-N curves}
The results from the simulations of both mode I and mixed mode I+II crack propagation demonstrate that the PF-CZM model can reasonably fit S-N curves within certain ranges. However, in the low-cycle fatigue regime, the model, based on the fatigue hypothesis used, struggles to accurately predict fatigue life. 
This results in shorter fatigue life predictions compared to experimental data (as evident in Fig.~\ref{fig:mode_I_results_LS4}e). A potential solution to this issue could involve exploring alternative fatigue damage hypotheses (e.g.,~\citep{BAKTHEER_2021_1}) or fatigue accumulation strategy (e.g.,~\citep{GOLAHMAR_2023}) that allow for better calibration of S-N curves in both low- and high-cycle regimes.
Furthermore,  for very high cycle fatigue, spanning millions of load cycles, the integration of this approach into a multiscale framework might become essential. Such integration will enable the establishment of connections to lower scales and thus the detection and triggering of defects at the nano-micro level and the understanding of their impact on fatigue behavior at the macro level~\cite{BENEDETTI_2018,  LOCASCIO_2021, PARRINELLO_2021}.

\paragraph{Paris law}
The analysis of crack growth rates, as shown in Fig.~\ref{fig:mode_I_results_LS4}f, indicates that the PF-CZM approach is capable of reasonably capturing the concept of the Paris law. 
Similar observations align with findings by other researchers in the field of phase field modeling, strengthening the credibility of this capability (e.g.,~\cite{Carrara_2020, selevs2021, ALESSI_2018, Yin_2023}). Further quantitative evaluation of the numerically calculated Paris law via the PF-CZM approach for fatigue crack propagation tests in concrete is still to be carried out in the future.

\paragraph{Hysteretic loops}
Due to the purely damage-based formulation and the fatigue damage hypothesis chosen, the PF-CZM approach presented is not inherently capable of reproducing the hysteretic loops in cyclic response. 
Achieving this would require consideration of elasto-plastic behavior in the cracked material, incorporating kinematic hardening, and adapting alternative fatigue damage hypotheses, that allow the accumulation of plastic deformations during the cycle, i.e. a ratchet mechanism, as demonstrated in previous studies (e.g.,~\cite{BAKTHEER_2021_1, ULLOA_2021}).

\paragraph{Fatigue creep curves}
The presented simulations of mode I and mixed mode I+II crack propagation clearly reveal the PF-CZM ability to reproduce the characteristic shape of fatigue creep curves. 
This includes accelerated growth in the initial and final phases, as well as linear and moderate growth in the intermediate phase (Fig.~\ref{fig:mixed_mode_results_LS4}).
A more accurate representation of fatigue creep curves can be achieved through the incorporation of the ratcheting mechanism, as shown in~\cite{BAKTHEER_2021_1, ULLOA_2021}.

\paragraph{Complex fatigue crack propagation}
As substantiated by numerous researchers, one of the principal strengths of the phase field approach lies in its proficiency for representing complex crack propagation phenomena. The presented outcomes across 2D and 3D examples subjected to fatigue loading further underscore this inherent capability of the PF-CZM methodology.

In summary, the critical evaluation highlights the potential of the PF-CZM approach and provides valuable insights into its current limitations. 
By exploring alternative fatigue damage hypotheses and incorporating aspects like elasto-plasticity and kinematic hardening, further refinements can enhance the predictive accuracy of the model and broaden its applicability to a wider range of fatigue scenarios.

\begin{table*}[!t]
\renewcommand*{\arraystretch}{1.25}
{ \centering
 \caption{Evaluation of PF-CZM approach in terms of key fatigue characteristics}
 \label{t:PFM_CZ_evaluation}
 \begingroup\setlength{\fboxsep}{0pt}
\colorbox{lightgray}{%
 \begin{tabular}{m{5.5cm} m{3.0cm}  m{7.8cm}}
    \toprule
    \textbf{Fatigue aspect} 
    & 
    \textbf{PF-CZM approach} 
    & 
    \textbf{Suggestion for improvement} 
    \\ \toprule 
    S-N curves 
    & 
    \multicolumn{1}{c}{\cmark|\xmark}
    &
    Considering other fatigue damage hypothesis or other damage accumulation strategy
    \\ \hline 
    Fatigue creep curves
    & 
    \multicolumn{1}{c}{\cmark}
    &
   Including the accumulation of plastic deformations , i.e. ratcheting mechanism
    \\ \hline 
    Hysteretic loops
    & 
    \multicolumn{1}{c}{\xmark}
    &
    Including plasticity with kinematic hardening as well as ratcheting mechanism
    \\ \hline 
    Paris law
    & 
    \multicolumn{1}{c}{\cmark}
    &
   -
    \\ \hline 
    Complex~fatigue crack propagation
    & 
    \multicolumn{1}{c}{\cmark}
    &
    -
    \\ \bottomrule 
 \end{tabular}%
}\endgroup
}
\mbox{}\\
{\small \cmark: can be accurately captured ;\;
\xmark: can not be captured;\;
\cmark|\xmark: can be captured, but with certain limitations.
}
\end{table*}

\section{Conclusions}

The presented PF-CZM approach, employed for modeling fatigue crack propagation in quasi-brittle materials like concrete, demonstrates remarkable validity when compared to experimental data. 
The critical evaluation, conducted across diverse loading scenarios and the examination of multiple fatigue characteristics, highlights its potential to realistically predict fatigue crack growth. 
This approach features reproduction of the classic S-shape of fatigue creep curves and effective modeling of complex crack propagation as observed in both 2D and 3D examples under fatigue loading.
Nevertheless, certain challenges remain. In particular, the fitting of S-N curves in the low fatigue cycle region and the lack of hysteretic loops in the PF-CZM formulation need further improvement. The remedies proposed in this study provide a way to improve the capabilities of the method.
In summary, the PF-CZM approach offers very promising capabilities for predicting fatigue crack growth in concrete. Addressing these limitations will provide a comprehensive and reliable tool for fatigue crack propagation analysis in quasi-brittle materials.

\end{document}